\documentclass[prd, aps, superscriptaddress, preprintnumbers, twocolumn, floatfix, nofootinbib]{revtex4-2}

\usepackage{amsfonts}
\usepackage{amsmath}
\usepackage{amssymb}
\usepackage{bm}
\usepackage{dcolumn}
\usepackage{graphicx}   
\usepackage[latin1]{inputenc}
\usepackage{latexsym}
\usepackage{rotating}
\usepackage{hyperref}
\usepackage{graphicx}
\usepackage{color}
\usepackage{xcolor}
\usepackage{ulem}
\usepackage{natbib}

\newcommand{\nn}{\nonumber \\}

\newcommand{\cO}{{\cal O}}

\newcommand{\gsim}{\mathrel{\hbox{\rlap{\lower.55ex \hbox {$\sim$}}
                   \kern-.3em \raise.4ex \hbox{$>$}}}}
\newcommand{\lsim}{\mathrel{\hbox{\rlap{\lower.55ex \hbox {$\sim$}}
                   \kern-.3em \raise.4ex \hbox{$<$}}}}

\newcommand\be{\begin{equation}}
\newcommand\ba{\begin{align}}
\newcommand\bas{\begin{align*}}
\newcommand\bt{\begin{table}}
\newcommand\bts{\begin{table*}}
\newcommand\bfig{\begin{figure}}
\newcommand\bfs{\begin{figure*}}
\newcommand\bi{\begin{itemize}}
\newcommand\ee{\end{equation}}
\newcommand\ea{\end{align}}
\newcommand\et{\end{table}}
\newcommand\ets{\end{table*}}
\newcommand\efig{\end{figure}}
\newcommand\efs{\end{figure*}}
\newcommand\ei{\end{itemize}}

\newcommand\tit{\textit}

\newcommand\bbf{\boldsymbol}

\hypersetup{colorlinks=true,
            breaklinks=true,
            pdfstartview=Fit,
            linkcolor=blue,
            citecolor=green,
            urlcolor=blue}
            
\begin{document}

\title{{Accretion onto Oscillating Cosmic String Loops}}

\author{Hao Jiao}
\email{hao.jiao@mail.mcgill.ca}
\affiliation{Department of Physics, McGill University, Montr\'{e}al, QC, H3A 2T8, Canada} 

\author{Bryce Cyr}
\email{bryce.cyr@manchester.ac.uk}
\affiliation{Jodrell Bank Centre for Astrophysics, School of Physics and Astronomy, The University of Manchester, Manchester M13 9PL, U.K.}

\author{Robert Brandenberger}
\email{rhb@physics.mcgill.ca}
\affiliation{Department of Physics, McGill University, Montr\'{e}al, QC, H3A 2T8, Canada} 

\date{\today}


\begin{abstract}

\noindent Cosmic string loops are non-linear density fluctuations which form in the early universe and could play an important role in explaining many phenomena which are in tension with the standard $\Lambda$CDM model. Hence, the details of the accretion process onto cosmic string loops should be studied in detail. Most previous works view loops as point masses and ignore the impact of a finite loop size. In this work, we utilize the Zel'dovich approximation to calculate the non-linear mass sourced by a static extended loop with a time-averaged density profile derived from the trajectory of the loop oscillation, and compare the result with what is obtained for a point-mass source. We find that the finite size of a loop mainly affects the evolution of turnaround shells during the early stages of accretion, converging to the point mass result after a critical redshift, $z^{({\rm II})/({\rm III})}_{\rm c}$. For $z>z^{({\rm II})/({\rm III})}_{\rm c}$, the total accreted mass surrounding a loop is suppressed relative to the point mass case and has a growth rate proportional to $(1+z)^{-3/2}$.
As an immediate extension, we also qualitatively analyse the accretion onto moving point masses and onto moving extended loops. In addition to the reduction in the nonlinear mass, the loop finite size also changes the shape of the turnaround surface at early stages of accretion.

\end{abstract}

\maketitle

~~~~

\section{Introduction} \label{ch1}
Cosmic strings are one-dimensional topological defects predicted in many particle physics theories beyond the Standard Model \cite{CS1,CS2,CS3,CS4}.  In theories which admit string solutions, a network of strings inevitably forms during a symmetry breaking phase transition in the early universe and persists to the present time \cite{Kibble1,Kibble2}.  At times $t$ sufficiently long after the phase transition, the distribution of strings consists of a network of {\it long} strings with mean curvature radius and separation comparable to the Hubble radius $t$, and a distribution of string loops which are created by the intersection of the long strings and are required to maintain the scaling behaviour of the long strings. Since cosmic strings are relativistic objects, long strings will typically be moving through the plasma of the early universe with relativistic speeds, generating non-linear overdensities in their wake \cite{wake1,wake2,wake3}. Besides, string loops act as seeds, generating non-linear density fluctuations at arbitrarily high redshifts. 

The gravitational effects of cosmic strings are given in terms of the string tension $\mu$ which is the mass of cosmic strings per unit length and usually expressed in dimensionless units as $G \mu$, where $G$ is Newton's gravitational constant.  Initially, it was conjectured that string loops could explain the origin of the large-scale structure in the universe \cite{early1,early2,early3}, which would require a value $G\mu \sim 10^{-6}$. Strings generate active and incoherent fluctuations, and thus, a model in which the density fluctuations are due completely to strings would not yield acoustic oscillations in the angular power spectrum of the CMB \cite{noacoustic1,noacoustic2}.  Since their oscillations have now been detected, we conclude that cosmic strings can only be a secondary source of fluctuations. The CMB data yields a bound of $G\mu \leq 10^{-7}$ \cite{CS-CMB-1,CS-CMB-2,CS-CMB-3}.  Although this bound rules out particle physics models producing cosmic strings at the high end of the scale of Grand Unification, there is a vast set of Beyond the Standard Model particle physics theories which predict strings with tensions lower than the abovementioned bound.  Such strings will be a subdominant contribution to the observed structure of the universe today, but since strings yield non-linear fluctuations at high redshifts in contrast to the standard $\Lambda$CDM model in which the number density of non-linear objects of a fixed mass decays exponentially with redshift, strings may play an important role at high redshifts. Thus, many puzzles in the $\Lambda$CDM model based on Gaussian primordial fluctuations can be solved by cosmic strings. 

Cosmic strings lead to distinct signatures in a wide range of observational windows (see e.g. \cite{RHBrev} for a short review). Long strings lead to line discontinuities in CMB anisotropy maps \cite{CS-CMBaniso-1,CS-CMBaniso-2} which can be searched for using various statistics designated to pick out the specific non-Gaussian signals (see e.g. \cite{CS-CMB-4,CS-CMB-5,CS-CMB-6,CS-CMB-7}). Long strings also give rise to patches in the CMB sky with extra polarization \cite{Holder1}, and to thin slices of extra absorption in high redshift 21-cm maps \cite{Holder2,wake-LSS}.  They also lead to planar overdensities of galaxies at higher redshifts \cite{Disrael}. Cross-correlation studies can also provide an interesting avenue to search for the signals of the long string distribution \cite{Hannah}.

Until recently there has been less work on the signatures of string loops. Since loops oscillate and slowly decay by emitting gravitational radiation,  bounds on $G\mu$ can be derived from bounds on the stochastic background of gravitational waves derived using pulsar timing array studies.  Data from the North American Nanohertz Observatory for Gravitational Waves (NANOGrav; \cite{NANOgrav1,NANOgrav2,CS-NANO-2}) give the tightest constraint on the cosmic string tension of $G\mu \lsim 10^{-10}$ \cite{CS-NANO-1,CS-NANO-2,CS-NANO-3} (the precise value, however, depends on assumptions about the distribution of string loops and is hence not as robust as the CMB limit mentioned above). In light of the recent detection of a stochastic gravitational wave background \cite{NANOGrav2023det}, cosmic strings have seen an exciting resurgence \cite{Ellis2020, Wang2023}, however, the spectral index seen by NANOGrav is hard to reconcile with the stable cosmic strings we discuss here \cite{NANOGrav2023}.
 
As first discussed in \cite{Jerome}, cosmic string loops generated in the early universe (before the time of recombination) provide non-linear seed density fluctuations which could seed the observed abundance of supermassive black holes. Recently, it was in fact shown that for superconducting string loops, all conditions for {\it Direct Collapse Black Hole} formation can be satisfied at high redshifts \cite{Bryce}. An update on the allowed parameter values was conducted in \cite{Bryce2023}, where it was found that superconducting strings can offer a rather convincing explanation to the origin of the anomalous radio synchrotron background \cite{Cyr2023RSB, Fixsen2009, Dowell2018, Singal2022, Cowie2023}. 

At any given time $t$, there exists a distribution of string loops with radius up to $\alpha t$, where $\alpha$ is a constant of the order of $10^{-1}$ (based on cosmic string network evolution simulations \cite{CSsimuls1,CSsimuls2,CSsimuls3,CSsimuls4,CSsimuls5,CSsimuls6,CSsimuls7,CSsimuls8,CSsimuls9,CSsimuls10}), therefore, string loops can potentially provide seeds for both intermediate-mass and super-massive black holes \cite{IMBH}. In addition, cosmic string loops are able to source the high-redshift galaxy candidates detected by JWST \cite{JH-JWST} (whose presence might again be hard to explain in the standard $\Lambda$CDM model).

Hence, it is important to study the evolution of the overdensities sourced by cosmic string loops.  Most previous works on string loop accretion treat loops as point masses \cite{CS1,CS2,CSpt2,movingCSpt1}. However, this assumption might not be a good one for large loops, whose scale is not negligible compared to the size of the turnaround shells. Loops formed in the matter-dominated era have radii comparable to the size of galaxies, and hence the extended nature of the loop mass distribution will significantly affect how they can accrete matter, resulting in a different mass function for large loops compared to what would be obtained in the point mass approximation. Besides, if we are interested in the structure seeded by loops at high redshifts, relevant e.g. to loop-seeded direct collapse black holes \cite{Bryce}, we need to consider specifically the early phase of accretion, in which case the turnaround shell might be smaller than the loop scale. In these cases, it is also unreasonable to view the loop as a point mass and the growth rate formalism must be revisited, which is the main aim of the work presented here.

Shlaer \tit{et. al.} discussed the accretion onto a finitely extended loop in \cite{CS-structure}, but they assumed that only shells outside the boundary of the loop can feel its gravity. We take a more general approach by considering the accretion of shells originating inside a given loop by making use of a time-averaged density profile induced by the rapid oscillations of the loop.

In the following section, we briefly review the properties of oscillating cosmic string loops and derive their density profile. We then discuss matter accretion onto static cosmic string loops both in the point mass approximation and by including the extended nature of the source in Section \ref{ch3}. We derive the non-linear mass sourced by the extended mass distribution of the loop analytically, making use of a small angle approximation, and verify that the result matches well with a more numerical approach which does not require this approximation. In Section \ref{ch4}, we analyze the accretion onto moving loops following a similar formalism to the one developed in Section \ref{ch3}.  We summarize our conclusions and discuss the results in Section \ref{ch5}.

Here we work in natural units with $c=k_{\rm B}=\hbar=1$. We also use the scale factor, $a(t)$, to describe the expansion of the homogeneous and isotropic universe, which is a monotonic function of time $t$.  We will alternate between using the scale factor or the redshift $1 + z = 1/a(t)$ to replace time. We normalize the scale factor to be equal to unity at the present time, $t_0$, i.e. $a(t_0)=1$, and we take the redshift of matter-radiation equality to be $1 + z_{\rm eq} = 3400$, i.e. $a(t_{\rm eq})=1/3400$.

\section{Cosmic String Loop Oscillations} \label{ch2}

We focus on cosmic string loops in this work, with the assumption of the \tit{one-scale} model \cite{one-scale-1,one-scale-2,one-scale-3}. In this framework, the radius of a cosmic string loop is a time-independent fraction $\alpha\sim\mathcal{O}(0.1)$ \cite{CSsimuls4} of the Hubble scale at its formation time $t_{\rm f}$, i.e. $R_0(t_{\rm f})\sim\alpha t_{\rm f}$. The total length of a loop is given by $L = \beta R_0$ where $\beta = 2\pi$ in the case of a perfectly circular loop. Simulations of cosmic strings \cite{CSsimuls1,CSsimuls2,CSsimuls3,CSsimuls4,CSsimuls5,CSsimuls6,CSsimuls7,CSsimuls8,CSsimuls9,CSsimuls10} show that loops can take many shapes \cite{CSsimuls1,CSsimuls2}, and we assume a fiducial value of $\beta=10$ for the purposes of our computations. These are typical values obtained by simulations based on the Nambu-Goto model \cite{CSsimuls1,CSsimuls2,CSsimuls3,CSsimuls4,CSsimuls5,CSsimuls6,CSsimuls7}, in which the cosmic strings are viewed as exact one-dimensional structures. This model predicts that loops oscillate at relativistic speeds at all times \cite{CS1,CS2}, which we make use of when deriving the time-averaged density profile below.

Cosmic string loops in general have a complicated spatial structure, oscillating about their center of mass in all three dimensions. To make the calculation more tractable, we make two approximations. First, we consider the mass distribution to be spherically symmetric. When considering time scales much larger than the loop oscillation time, this approximation appears to be reasonable. Second, we determine the radial density profile by studying the time-averaged density for a particular loop configuration, namely a circular loop, and assume that this density profile is valid for all loops.

As a starting point, the total energy of an oscillating loop with an instantaneous radius $R(t)$ is given by
\begin{align}
E(R)&=\beta\mu R(t)\gamma(R(t))=\beta\mu R(t)\left(1-\dot R(t)^2\right)^{-1/2},
\end{align}
where $\gamma(R(t))$ is the Lorentz factor of the loop. The total energy of a loop should be constant when ignoring its decay. Then, we can obtain the expression of the instantaneous radius 
\ba
R(t)=R_0\cos\left(\frac{t}{R_0}\right),
\end{align}
where $R_0$ is the maximum radius of the oscillating loop, and the total energy at any time is 
$E(R)\equiv\beta\mu R_0=M_{\rm loop}$. 
We note that $R_0$ is the usual value of the loop radius commonly used in the literature. 

The oscillation timescale ($\sim$$\,R_0$) is always much smaller than the collapse timescale of matter shells around the loop  ($\sim$$\,t_{\rm H}$). Therefore, when considering the accretion onto a cosmic string loop, we should take into account these oscillations by viewing the loop as an extended source. We can then derive a time-averaged density profile to take this effect into account
\ba
\bar\rho(r)&=\frac{2}{\pi R_0}\int_0^{\pi R_0/2}dt\, \rho(r,t)\nn
&=\frac{\beta\mu}{2\pi^2 r^2 \sqrt{1-(r/R_0)^2}}, \label{eqrho}
\end{align}
where $\rho(r,t)=\frac{1}{4\pi r^2}E(R(t))\delta(r-R(t))$ is the instantaneous density profile of the oscillating loop with the assumption of spherical symmetry.

\section{Accretion onto static oscillating loops} \label{ch3}
To begin, we consider the standard accretion calculation viewing static loops as point masses, and then generalize to include the effects of loop oscillations. Later (in Sec.~\ref{ch4}), we further include finite loop velocity effects. Note that since the mean separation of cosmic string loops is much greater than the region affected by the loop gravity, we don't consider the influence of other loops on the accretion process. This accretion geometry is spherical since the time-averaged density profile of oscillating loops is spherical.

We use the Zel'dovich approximation \cite{Zel'dovich} to study the evolution of a shell around a cosmic string loop with physical height from the centre of loop given by
\be
h(q,t)=a(t)\left(q-\psi(q,t)\right),
\ee
where $a(t)=(t/t_0)^{2/3}$ is the scale factor in the matter-dominated era, $q$ is the initial comoving radius of the shell, and $\psi$ is the comoving displacement of the shell due to the loop. Note that the Zel'dovich approximation is only valid before the shell turns around ($\dot h(q,t)=0$). We can use this method to calculate the turnaround radius and then derive the non-linear mass of the loop-seeded overdensity, which we define to be the total mass inside this turnaround shell as the matter enclosed will collapse and virialize within roughly one Hubble time \cite{virial}. 

The equation of motion (EOM) for the displacement of a shell with height $h(q,t)$ to first order in $\psi$ is
\ba
\ddot\psi+\frac{4}{3}t^{-1} \dot\psi-\frac23 t^{-2}\psi&=\frac{GM(aq)}{q^2}\left(\frac{t_0}{t}\right)^2, \label{eqEOM}
\end{align}
where $M(aq)$ is the mass of the source enclosed within the physical initial coordinate $aq$. Note that since the displacement $\psi$ is small relative to the initial comoving coordinate $q$ before the shell turns around, we make use of the Born approximation, which allows us to ignore the impact of the perturbation $\psi$ on the enclosed mass. We derive this EOM and discuss more details in Appendix A. As a conservative estimate, we take the following initial conditions:
\ba
\psi(t_{\rm i})=\dot\psi(t_{\rm i})=0.
\end{align}
Here, $t_{\rm i}$ is the time when accretion begins. For loops that formed before $t_{\rm eq}$, accretion begins at $t_{\rm eq}$, i.e. $t_{\rm i}=t_{\rm eq}$ since the growth of perturbations is stifled at times earlier than this. For loop formation times $t_{\rm f}>t_{\rm eq}$, the accretion naturally begins at $t_{\rm i}=t_{\rm f}$.

\subsection{Point Mass Approximation} \label{ch3-1}
The accretion onto a loop approximated as a point source has been studied in previous works \cite{CS2,CSpt2}. We review the basics of the computation here, and make direct comparisons to the extended moving source in the following subsections. 

In the point source approximation, the enclosed mass is always the total mass of the loop, i.e. $M(aq)=\beta\mu R_0$, and Eq.~\eqref{eqEOM} becomes
\ba
\ddot\psi+\frac{4}{3}t^{-1} \dot\psi-\frac23 t^{-2}\psi&=\frac{\beta G\mu R_0}{q^2}\left(\frac{t_0}{t}\right)^2.
\end{align}
This can easily be solved, yielding the solution
\ba
\psi&=\frac32\frac{\beta G\mu R_0 t_0^2}{q^2}\left[\frac35\left(\frac{t}{t_{\rm i}}\right)^{2/3}-1+\frac25\left(\frac{t}{t_{\rm i}}\right)^{-1}\right]\nn
&\approx \frac{9}{10}\frac{\beta G\mu R_0 t_0^2}{q^2}\left(\frac{t}{t_{\rm i}}\right)^{2/3}.
\end{align}
For $t\gg t_{\rm i}$, the leading term is sufficient to describe the accretion process. A shell will turn around when $\dot h(q,t)=0$, allowing us to derive the comoving turnaround radius $q_{\rm nl}$,
\ba
q_{\rm nl}=\left(\frac95 \beta G\mu R_0 t_0^2\right)^{1/3}\left(\frac{1+z_{\rm i}}{1+z}\right)^{1/3},
\end{align}
where $z_{\rm i}$ is the redshift corresponding to the initial accretion time $t_{\rm i}$. The mass inside of a shell that turns around at time $t\gg t_{\rm i}$ is roughly
\ba
M_{\rm nl}^{({\rm pt})}=\frac{4\pi}{3}\rho^{\rm c}_{\rm bg}q_{\rm nl}^3=\frac25\beta\mu R_0 \left(\frac{1+z_{\rm i}}{1+z}\right).
\end{align}
Here, we note that $\rho_{\rm bg}^{\rm c} = 3 H_0^2/8\pi G$ is the comoving background energy density, related to the physical density during the matter epoch through $\rho_{\rm bg}^{\rm p} = \rho_{\rm bg}^{\rm c} (1+z)^3$. Therefore, the overdensity seeded by the point-source loop grows linearly in the scale factor, $a(t)$.

For loops that form before $t_{\rm eq}$, i.e. $R_0\leq \alpha t_{\rm eq}$, the initial accretion begins at a redshift of $1+z_{\rm eq}\simeq 3400$, so the corresponding nonlinear mass is
\ba
\left.M_{\rm nl}^{({\rm pt})}\right|_{\rm rad}&=\frac25\beta\mu R_0 \left(\frac{1+z_{\rm eq}}{1+z}\right)\nn
&\approx 6.1\times10^{11}M_\odot\cdot\left(\frac{G\mu}{10^{-10}}\right)\left(\frac{R_0}{t_{\rm eq}}\right)\left(1+z\right)^{-1}.
\end{align}
For loops forming in the matter era, the initial accretion time is approximately the formation time of the loop, i.e. $z_{\rm i}=z_{\rm f}=z(t_{\rm f})$. The formation time $t_{\rm f}$ is a function of the loop radius $R_0$
\ba
t_{\rm f}&=R_0/\alpha,
\end{align}
so we can easily find the relationship between loop radius and the redshift corresponding to $t_{\rm f}$,
\ba
1+z_{\rm f}&=
\left\{\begin{array}{ll}
(\alpha t_{\rm eq}/R_0)^{1/2}(t_0/t_{\rm eq})^{2/3} & (R_0< \alpha t_{\rm eq})\\
(\alpha t_0/R_0)^{2/3} & (R_0\geq \alpha t_{\rm eq})
\end{array}\right..
\end{align}
Therefore, the non-linear mass for loops that form during the matter era is modified, which becomes
\ba
\left.M_{\rm nl}^{({\rm pt})}\right|_{\rm mat}&=\frac25\alpha^{2/3}\beta t_{\rm eq}^{2/3}\mu R_0^{1/3}\left(\frac{1+z_{\rm eq}}{1+z}\right)\nn
&\approx 1.3\times10^{11}M_\odot\nn
&\ \ \ \ \ \times \left(\frac{G\mu}{10^{-10}}\right)\left(\frac{R_0}{t_{\rm eq}}\right)^{1/3}\left(1+z\right)^{-1}.
\end{align}

\subsection{Accretion onto an oscillating loop}  \label{ch3-2}

\begin{figure*}
	\centering
	\includegraphics[scale=0.6]{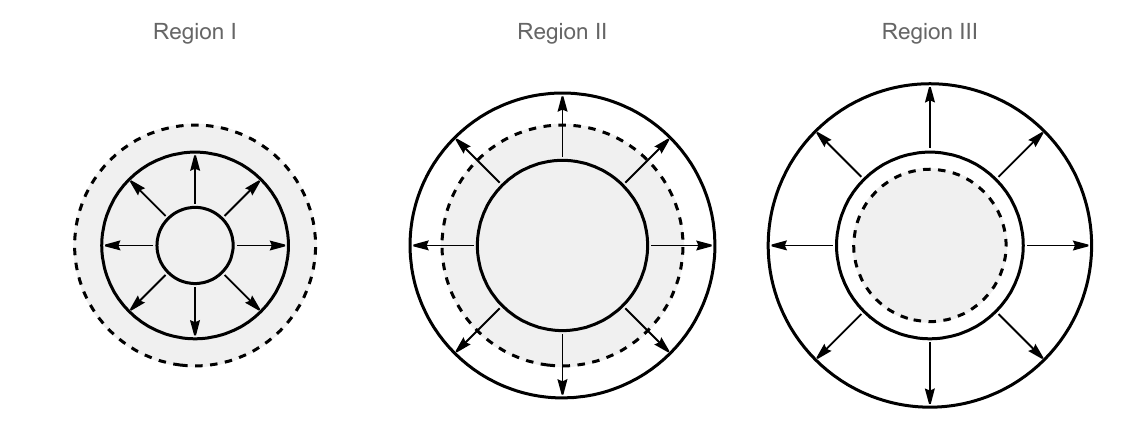}
	\caption{An illustration of the evolution of shells in Regions I, II, and III. The dashed circle corresponds to the loop radius $R_0$, while solid circles show the initial (inner circle) and the turnaround (outer circle) positions of a given test shell. The point mass approximation is valid only for scenarios described by Region III (the right panel). Note that these shells are shown in physical coordinates so the loop radius is constant.}
	\label{fig3-0}
\end{figure*}

In this subsection, we consider the accretion onto an oscillating loop. From the time-averaged density profile in Eq.~\eqref{eqrho}, we see that the enclosed mass at a given distance $aq$ from the centre of the loop is 
\ba
M(aq)&=\int_0^{aq} dr\,4\pi r^2\bar\rho(r)\nn
&=\left\{\begin{array}{ll}
\frac2\pi \beta\mu R_0\arcsin\left(\frac{aq}{R_0}\right)\ &(aq<R_0) \\
\beta\mu R_0\ &(aq\geq R_0)
\end{array}\right. .  \label{eqenclosedmass}
\end{align}

For shells inside the loop ($aq<R_0$), the enclosed mass is proportional to $\arcsin(aq/R_0)$, which will reduce the rate of accretion when compared to the point mass case. To derive some analytic expressions, we make use of the small angle approximation $\arcsin(x) \approx x$. Strictly speaking, this approximation is only valid for mass shells close to the centre of the loop, i.e. $aq\ll R_0$. We discuss this in more detail in the following subsections, where we present the exact numerical solution (without the small angle approximation). A comparison between the non-linear mass derived with and without this approximation can be seen in Fig.~\ref{fig3-2}. With this, the radial mass profile simplifies greatly
\ba
M(aq)\approx \frac2\pi \beta\mu aq. \label{eqapprox1}
\end{align}

To calculate the accretion onto an extended source, it is helpful to discuss the evolution of the turnaround shells in three separate regimes:
\bi
\item \tit{Region I}: Shells here originate and turn around while they are inside the boundary of the cosmic string loop ($a(t_{\rm ta})q\leq R_0$).
\item \tit{Region II}: Shells originate inside of the loop, but are dragged outside of it by Hubble expansion before they turn around, i.e. $a(t_{\rm i})q\leq R_0< a(t_{\rm ta})q$.
\item \tit{Region III}: Shells in this region evolve outside of the loop ($a(t_{\rm i})q> R_0$) at all times before turnaround. The point mass approximation is valid for these shells alone.
\ei

An illustration of these three regimes can be seen in Fig.~\ref{fig3-0}. The overall evolution of mass shells in each of these regimes proceeds differently, as the structure of $M(aq)$ changes. 

We assume that shells do not cross each other before turnaround, meaning that inner shells will collapse earlier than outer shells. The growth of non-linear mass around a cosmic string loop thus proceeds in three phases. First, shells in Region I collapse, followed by those in Region II, and finally in Region III.

\subsubsection{Region I} \label{ch3-2-1}
In this section, we study the evolution of shells in Region I, which will turn around and collapse before they reach the boundary of the cosmic string loop. Making use of the small angle approximation for the enclosed mass (Eq.~\eqref{eqapprox1}), the equation of motion for $\psi$ is given by
\ba
\ddot\psi+\frac{4}{3}t^{-1} \dot\psi-\frac23 t^{-2}\psi&=\frac2\pi\frac{\beta G\mu}{q}\left(\frac{t_0}{t}\right)^{4/3},\label{eq8}
\end{align}
with the two initial conditions
\ba
\psi(t_{\rm i})=\dot\psi(t_{\rm i})=0.
\end{align}
The solution in this region is
\ba
\psi^{({\rm I})}(t)& = \frac6{5\pi}\frac{\beta G\mu}{q}t_0^{4/3}t_{\rm i}^{2/3}\nn
&\ \ \ \ \times \left[\left(\frac{t}{t_{\rm i}}\right)^{2/3}\ln\left(\frac{t}{t_{\rm i}}\right)-\frac35\left(\frac{t}{t_{\rm i}}\right)^{2/3}+\frac35\left(\frac{t}{t_{\rm i}}\right)^{-1}\right] \nonumber \\
&\approx \frac6{5\pi}\frac{\beta G\mu}{q}t_0^{4/3}t^{2/3}. \label{eq:psiApprox}
\end{align}
Again, the approximation in the final line is valid for $t\gg t_{\rm i}$ where we also set the logarithm term to unity since it increases much slower than the power law. This introduces an $\mathcal{O}(1)$ uncertainty on $\psi^{({\rm I})}$, though we will see in the following subsection that this approximation generally reproduces the results we see when performing an exact numerical calculation provided that we are not too near the initial accretion time.

The comoving turnaround radius is found by solving 
\ba
\dot h&=\dot a\bigl(q_{\rm nl}^{({\rm I})}-\psi^{({\rm I})}\bigr)-a\dot\psi^{({\rm I})}=0,
\end{align}
and is given by 
\ba
q_{\rm nl}^{({\rm I})}(z)&=\bigg\{\frac{18\beta G\mu}{5\pi}t_{0}^2(1+z)^{-1} \nn
& \hspace{8mm}\times\bigg[\ln \left(\frac{1+z_{\rm i}}{1+z}\right)+\frac{1}{10}-\frac{1}{10}\left(\frac{1+z_{\rm i}}{1+z}\right)^{-5/2}\bigg]\bigg\}^{1/2} \nonumber \\
&\approx\left[\frac{18}{5\pi}\beta G\mu\,t_0^2(1+z)^{-1}\right]^{1/2}. \label{eq:qIapprox}
\end{align}

Here, we again take only the leading term and drop the logarithm in the last line. The non-linear mass corresponding to this turnaround radius is
\ba
M_{\rm nl}^{({\rm I})}&=\frac{4\pi}{3}\rho_{\rm bg}^{\rm c} q_{\rm nl}^3 \nonumber \\
&=7.6\times10^8\,M_\odot\cdot\left(\frac{G\mu}{10^{-10}}\right)^{3/2}(1+z)^{-3/2}\nn
&\ \ \ \ \ \times\left[\ln\left(\frac{1+z_{\rm i}}{1+z}\right)+\frac{1}{10}-\frac{1}{10}\left(\frac{1+z_{\rm i}}{1+z}\right)^{-5/2}\right]^{3/2} \label{eq-MnlI-1}\\
&\approx 7.6\times10^{8}M_\odot\cdot \left(\frac{G\mu}{10^{-10}}\right)^{3/2}\left(1+z\right)^{-3/2}.\label{eq-MnlI}
\end{align}
The last line expression gives the leading-order growth rate for $z \gg z_{\rm i}$. Immediately we can see that this non-linear mass no longer grows linearly ($\propto a$), but instead, we have $M_{\rm nl}^{({\rm I})}\propto a^{3/2}$. Naively, the accretion onto an extended source seems to increase faster than onto a point mass. This is because as a shell expands, the enclosed mass increases and the source term for $\psi$ becomes stronger, causing accretion to speed up. Similarly, the dependence of the nonlinear mass on the string tension changes from $G\mu$ to $(G\mu)^{3/2}$ as a result of the gradual increase of the enclosed mass. However, as the amplitude of the source term here is always $M(h) \leq \beta \mu R_0 = M_{\rm loop}$, the nonlinear mass in Eq.~\eqref{eq-MnlI} is smaller than what is obtained in the point mass case until a given shell reaches $R_0$ and $M_{\rm nl}$ returns to the less steep scaling ($\propto a$) in Region III.

An interesting result is that the non-linear mass is independent of the loop radius $R_0$ in this region. This is simply a consequence of the small angle approximation we utilized in Eq.~\eqref{eqapprox1}. Again, we reiterate that Eq.~\eqref{eq-MnlI} is only valid for ``steady state" accretion in Region I ($z \gg z_{\rm i}$), and as a result $M_{\rm nl}^{({\rm I})}(z_{\rm i}) \neq 0$. To build intuition, our approximations capture only the leading-order power-law growth in the steady state. The full solution for the growth in Region I (utilizing the small angle approximation, but keeping all subleading terms) is shown in Eq.~\eqref{eq-MnlI-1}. The comparison between the leading-order behaviour and this full solution can be seen by the orange and red dashed lines in Fig.~\ref{fig3-2}.

\subsubsection{Region II} \label{ch3-2-2}
We now move on to discussing the accretion of a shell in Region II, which originates inside the boundary of the loop, but turns around after crossing the loop radius. 

We mark the onset of Region II accretion as the redshift at which shells begin turning around at the boundary of the loop radius $R_0$. Since $\psi(t_{\rm ta}) \sim q/2$ at the turnaround time $t_{\rm ta}$ (details in Appendix \ref{App-A}), the critical redshift between Regions I and II, $z_{\rm c}^{({\rm I})/({\rm II})}$, can be derived by requiring $R_0 = a q^{({\rm I})}_{\rm nl}/2$:
\ba \label{eq:zI-II}
\bigg(\frac{1+z_{\rm c}^{({\rm I})/({\rm II})}}{1+z_{\rm eq}}\bigg)&=\left[ \frac{9}{10\pi} \beta G\mu \left(\frac{t_{\rm eq}}{R_0}\right)^2 \right]^{1/3}.
\end{align}
This is the redshift at which a loop with a given $G\mu$ and $R_{0}$ will transition from Region I to Region II accretion. We illustrate this critical redshift by the dashed lines for two typical values of the string tension $G\mu$ in Fig.~\ref{fig3-1}.

A shell in Region II evolves in the same way as in Region I before it crosses the loop boundary and then the source term stops growing. When the physical radius of the shell is greater than the loop radius, the EOM \eqref{eq8} is no longer valid and must be modified to reflect this fact.

We need to determine the time that the height of a given mass shell crosses the loop radius, $t_{\rm cross}$. To do this, we simply equate the height of a shell to $R_0$ and find
\ba
h_{\rm cross}&=a(t_{\rm cross})(q-\psi(q,t_{\rm cross}))\simeq a(t_{\rm cross})q=R_0, \nonumber \\
\Rightarrow t_{\rm cross}&\simeq\left(R_0/q\right)^{3/2}t_0.\label{eq:tcross}
\end{align}
Here we posit that $q \gg \psi(q,t_{\rm cross})$, which is a basic assumption of the Zel'dovich approximation and is valid for all shells in Region II as they turn around after crossing the loop.

To compute the Region II growth, we use the results from Region I as initial conditions at $t_{\rm cross}$. This yields
\ba
\psi(t_{\rm cross})&=\frac6{5\pi}\frac{\beta G\mu}{q}t_0^{2}\left(\frac{t_{\rm cross}}{t_0}\right)^{2/3},\label{eq-iniII-1}\\
\dot\psi(t_{\rm cross})&=\frac4{5\pi}\frac{\beta G\mu}{q}t_0\left(\frac{t_{\rm cross}}{t_0}\right)^{-1/3}.
\label{eq-iniII-2}
\end{align}
For $t>t_{\rm cross}$, we can once again view the loop as a point source with mass $M_{\rm loop}=\beta\mu R_0$, and the EOM of $\psi$ becomes 
\ba
\ddot\psi+\frac{4}{3}t^{-1} \dot\psi-\frac23 t^{-2}\psi&=\frac2\pi\frac{\beta G\mu R_0}{q^2}\left(\frac{t_0}{t}\right)^2.\label{EOM-II}
\end{align}

\bfig
	\centering
	\includegraphics[scale=0.55]{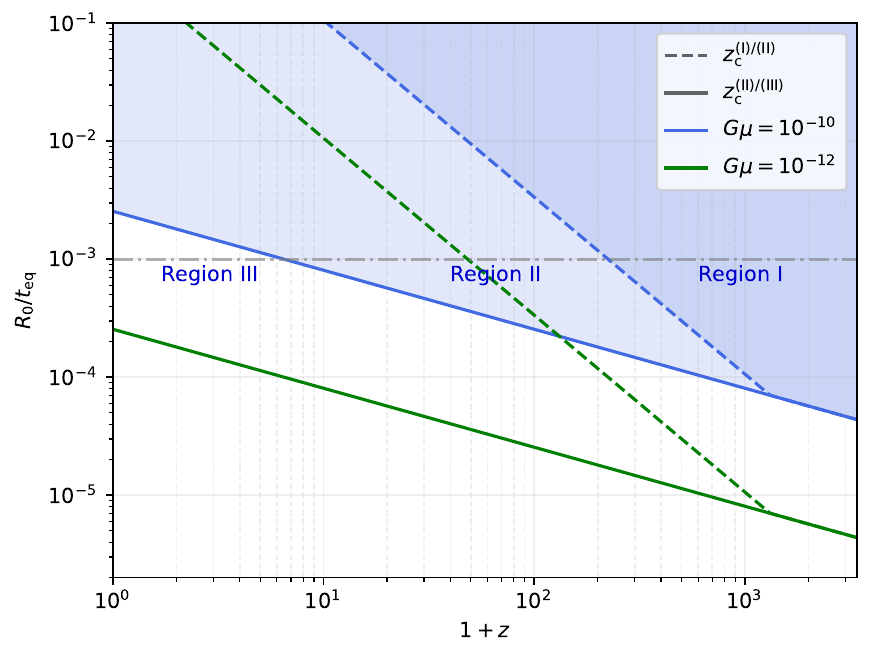}
	\caption{An illustration of the accretion regimes as a function of redshift for a loop with a given radius ($R_0/t_{\rm eq}$), using two typical values for the string tension. The dashed-dotted line corresponds to a loop with $R_0 = 10^{-3} t_{\rm eq}$, a benchmark that we discuss in section \ref{ch3-3}. For the $G\mu = 10^{-10}$ case (blue lines), the darker blue region corresponds to Region I accretion, with the lighter blue shows when Region II is effective, and the unshaded region is where the point mass approximation is valid (Region III). The $G\mu = 10^{-12}$ case (green lines without shaded region) shows how these regions change with decreasing string tensions.}
	\label{fig3-1}
\efig

We retain the factor of $2/\pi$ in the source term from the small angle approximation to ensure continuity of $\psi$ between Regions I and II. The corresponding solution of the above equation is
\ba
\psi^{({\rm II})}&=\frac3\pi\frac{\beta G\mu}{q}t_0^2\left(\frac{t}{t_0}\right)^{2/3}\left[1 - \frac{R_0}{a(t)q} + \frac{2}{5} \left(\frac{R_0}{a(t)q}\right)^{5/2} \right] \nonumber \\
&\approx\frac3\pi\frac{\beta G\mu}{q}t_0^2\left(\frac{t}{t_0}\right)^{2/3}.
\end{align}

Once a shell has crossed the loop radius, we have $t > t_{\rm cross}$ (equivalently $a(t)q > R_0$). As a result, we only keep the leading term in the second line, which corresponds to the steady state solution in Region II (valid for $t \gg t_{\rm cross}$). The non-linear comoving radius at a given redshift $z$ is then given by
\ba
q_{\rm nl}^{({\rm II})}&=\left[\frac6\pi\beta G\mu\,t_0^2\left(1+z\right)^{-1}\right]^{1/2}, \label{eq29}
\end{align}
and we can approximate the non-linear mass as\footnote{Note that the the matching condition is not satisfied ($M_{\rm nl}^{{\rm (I)}}(z_{\rm c}^{({\rm I})/({\rm II})}) \neq M_{\rm nl}^{{\rm (II)}}(z_{\rm c}^{({\rm I})/({\rm II})})$). This is because the non-linear masses we derive are valid only for their respective steady state regimes. The exact numerical solution contains no such discontinuities.}
\ba
M_{\rm nl}^{({\rm II})}\approx1.6\times10^9 M_\odot \left(\frac{G\mu}{10^{-10}}\right)^{3/2}\left(1+z\right)^{-3/2} \label{eq-MnlII}.
\end{align}
We find that the time-dependence of the non-linear mass in Region II is the same as in Region I if we only consider their respective steady state regimes. This is because the coefficient of the dominant term of $\psi^{\rm (II)}$ inherits the same $q-$dependence as $\psi^{(I)}$ from the initial conditions in Eqs. \eqref{eq-iniII-1} and \eqref{eq-iniII-2}. Furthermore, there is only roughly a factor of $2$ difference between non-linear mass in Regions I and II, due to the fact that the two turnaround radii are related by a similar factor. Thus, ``crossing the loop radius" seems to have little effect on the evolution of a turnaround shell that originates inside of the loop. It is therefore a reasonable approximation to not distinguish between Regions I and II in our analytic treatment of the growth of non-linear mass.

\subsubsection{Region III} \label{ch3-2-3}
The shells in Region III never enter the loop before turnaround, so the point mass treatment is justified in this regime. As shown in section \ref{ch3-1}, the overdensity grows linearly in the scale factor.

The collapse of shells occurs sequentially from Region I, to Region II, and finally into Region III, so we need to determine at which redshift shells with initial heights $h(t_i)=a(t_{\rm i}) q \geq R_0$ begin turning around and contributing to the non-linear mass. This is the time that the accretion enters into Region III. 

A simple way to determine this redshift is to compute the time when the total non-linear mass is equal to the background mass inside the loop radius at the initial accretion time ($t_{\rm i}$). The critical redshift ($z_{\rm c}^{({\rm II})/({\rm III})}$) at which shells in Region III begin turning around is given by
\ba
M_{\rm nl}^{({\rm II})}\left(z_{\rm c}^{({\rm II})/({\rm III})}\right)&=\frac{4\pi}{3}R_0^3 \rho_{\rm bg}^{\rm p}(t_{\rm i}),\\
\Rightarrow\ \ \bigg(\frac{1+z_{\rm c}^{({\rm II})/({\rm III})}}{1+z_{\rm eq}}\bigg)&=\frac{6\beta}{\pi}G\mu\bigg(\frac{t_{\rm eq}}{R_0}\bigg)^2\bigg(\frac{1+z_{\rm i}}{1+z_{\rm eq}}\bigg)^{-2}. \label{eq-zc}
\end{align}
In Fig.~\ref{fig3-1}, we show this critical redshift as a function of loop radius in units of $t_{\rm eq}$ by the solid lines (note that $R_0 \simeq 0.1 t_{\rm eq}$ is roughly the size of loops produced at matter-radiation equality).

From this figure, we can see that only loops with radii in a certain range, will exhibit Region III growth. Taking $G\mu=10^{-10}$ as a benchmark,  loops with radii larger than $R_0 > 2.5\times10^{-3}t_{\rm eq}$ will never enter Region III, whereas smaller loops with $R_0\leq 4.4\times10^{-5}t_{\rm eq}$ will be in Region III soon after matter radiation equality. The presence of Region II accretion appears to ``emerge" at $z\simeq 1200$ independent of loop size. This effect is artificial and is due to the fact that Eq.~\eqref{eq:zI-II} was derived using the leading-order analytical solution in Region I, which underestimates the non-linear mass at the boundary of Regions I and II. This further propagates the relative suppression into Eq.~\eqref{eq:zI-II}. This mismatch can also be seen by comparing the red and orange lines in Fig.~\ref{fig3-2} at the boundary of Regions I and II. The numerical and full analytic results always proceed sequentially through Regions I-III, but do not have compact analytic forms.

For loops that form after $t_{\rm eq}$, i.e. $R> 0.1t_{\rm eq}$, Region III is never entered and the point mass approximation is thus never justified. Therefore, we only need to consider the Region III accretion of loops forming in the radiation phase and do not distinguish between the initial accretion time $t_i$ and matter-radiation equality $t_{\rm eq}$ in what follows. 

For $z<z_{\rm c}^{({\rm II})/({\rm III})}$, the non-linear mass grows linearly ($\propto a(t)$) as was derived in the point mass case, so we have
\ba
M_{\rm nl}^{({\rm III})}(z,R)&=M_{\rm nl}^{({\rm II})}(z_{\rm c}^{({\rm II})/({\rm III})})\left(\frac{1+z_{\rm c}^{({\rm II})/({\rm III})}}{1+z}\right) \nonumber\\
&=\frac{4}{3\pi}\beta \mu R_0 \left(\frac{1+z_{\rm eq}}{1+z}\right). \label{eq-MnlIII}
\end{align}

\subsection{Comparison to the numerical solution} \label{ch3-3}
In the previous subsection, we derived some leading-order approximations to the growth rate of non-linear mass around a (static) oscillating cosmic string loop. Here, we investigate the accuracy of these approximations. It is possible to solve the EOM for $\psi$ inside of the loop without the small angle approximation ($\arcsin(x)\approx x$) in a fully analytic way. The solution, however, is sufficiently complicated that the computation of the turnaround radius $q_{\rm nl}$ and non-linear mass $M_{\rm nl}$ is only possible numerically. The details of this calculation can be found in Appendix~\ref{App-B}.

\bfig
	\centering
	\includegraphics[scale=0.55]{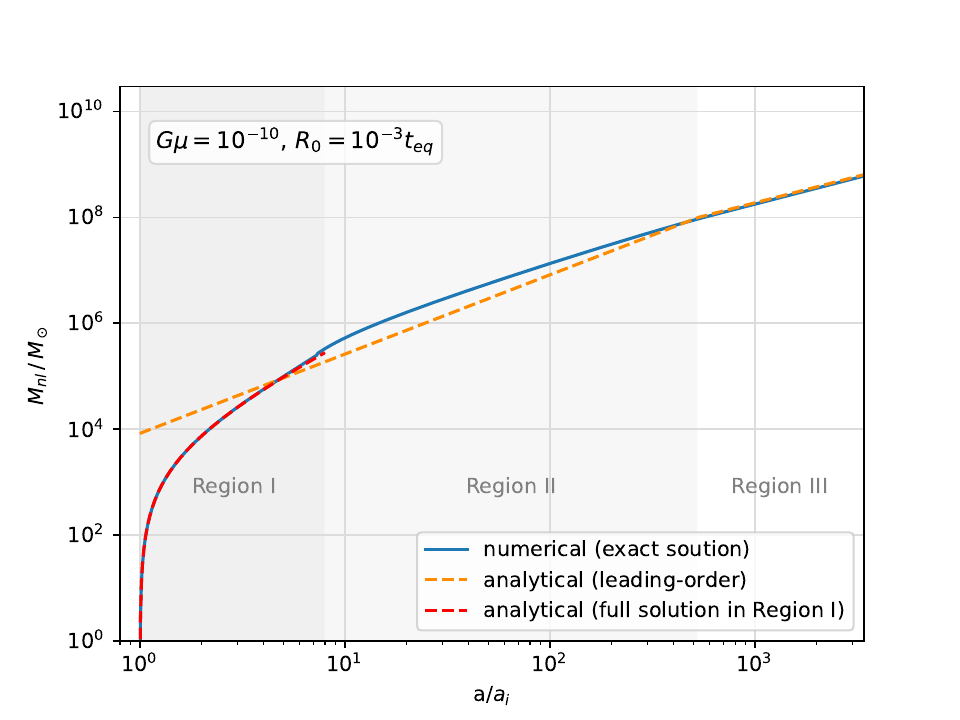}
	\caption{Comparing the non-linear mass as a function of $a/a_{\rm i}$ for the cases with (dashed) and without (solid) the small angle approximation ($\arcsin (x)\approx x$).  Here we set $G\mu=10^{-10}$ and the loop radius $R_0=10^{-3} t_{\rm eq}$. The solid blue line shows the exact numerical solution, while the orange dashed line corresponds to the leading-order non-linear masses derived in Eqs.~\eqref{eq-MnlII} and \eqref{eq-MnlIII}. The red dashed line shows the full expression of $M_{\rm nl}^{({\rm I})}$ in Eq.~\eqref{eq-MnlI-1}, which is only valid in Region I.}
	\label{fig3-2}
\efig

We compare the non-linear mass $M_{\rm nl}$ computed with and without the small angle approximation in Fig.~\ref{fig3-2}. Here we take as benchmark a loop with radius $R_0=10^{-3}t_{\rm eq}$ and string tension $G\mu=10^{-10}$, which is a rather generic and instructive case as it undergoes accretion in Regions I-III during the matter era (see the dash-dotted line in Fig.~\ref{fig3-1}).

The solid blue line in Fig.~\ref{fig3-2} shows the exact numerical solution of the non-linear mass without any approximations (see Appendix \ref{App-B}). The orange dashed line corresponds to the non-linear masses we derived in the previous subsection, namely Eqs.~\eqref{eq-MnlII} and \eqref{eq-MnlIII}\footnote{Recall that we approximate $M_{\rm nl}^{({\rm I})} \approx M_{\rm nl}^{({\rm II})}$ in the previous section.}. Finally, the red dashed line is the full solution to the Region I accretion using the small angle approximation, but keeping terms which are subdominant at $t\gg t_{\rm i}$ in Eq.~\eqref{eq-MnlI-1}. Regions I, II and III are shown by gray, light gray and white regions with corresponding labels allowing us to compare the evolution of the non-linear mass in the three regions separately.

The loops in our benchmark model form during the radiation-dominated era, and begin to accrete matter at $t_{\rm i}=t_{\rm eq}$. Thus, the redshift range of interest is $1\geq a/a_{\rm i}\gtrsim 3400$. For loops forming in the matter-dominated era (which are significantly larger with $R_0 \gtrsim 0.1 t_{\rm eq}$), the total accretion time will be shorter.

From this figure, we find that the full analytic non-linear mass in Eq.~\eqref{eq-MnlI-1} fits the numerical solution very well in Region I as the dashed red line almost overlaps the solid blue line. In Region II, solving the EOM analytically requires that we only keep the leading term in the solution of $\psi$ for the initial conditions in Eqs.~\eqref{eq-iniII-1} and \eqref{eq-iniII-2} and hence the approximate non-linear mass in Region I and II (the orange dashed line) is slightly different from the numerical solution. The maximum discrepancy between the ``steady state'' analytical nonlinear mass and the numerical result is roughly a factor of 2.5 in Region II. As we enter Region III, the numerical and approximate solutions converge once more. This also verifies the validity of our approximations as the nonlinear mass in Region III (Eq.~\eqref{eq-MnlIII}) is derived from the approximate solution in Region II, which only retained the leading term.

\bfig
	\centering
	\includegraphics[scale=0.55]{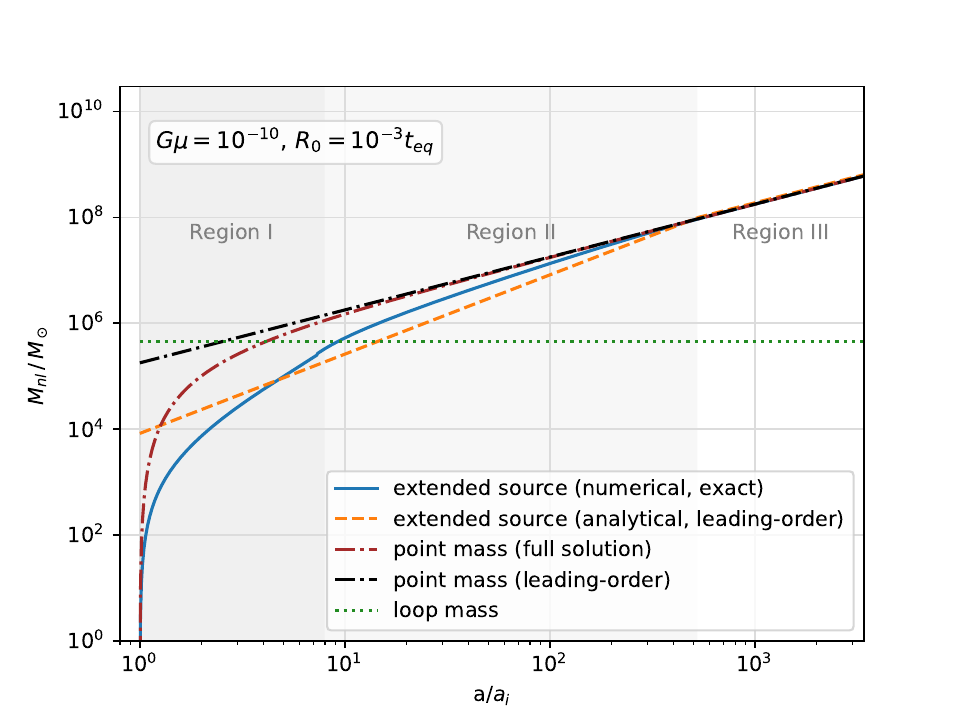}
	\caption{A comparison of the non-linear mass seeded by the loop when viewed as an extended source versus a point mass. The solid blue and dashed orange lines are the numerical (exact) and analytical (leading-order) accreted masses of the extended loop respectively. The red and black dash-dotted lines show the exact and leading-order masses of the overdensity sourced by loops in the point-mass case, where we can clearly see that they overestimate the Region I growth rate. The dotted green line corresponds to the rest-mass of a loop with radius $R_0=10^{-3}t_{\rm eq}$. }
	\label{fig3-3}
\efig

In Fig.~\ref{fig3-3}, we compare the full and leading-order non-linear masses seeded by a loop with $R_0=10^{-3}t_{\rm eq}$, considering both the point mass and extended source formalisms. For the extended source case, we illustrate the exact numerical (solid blue) and leading-order analytical (dashed orange) non-linear mass with the same parameters in Fig.~\ref{fig3-2}, while the accretion by a point-mass loop is shown by the dash-dotted lines. We find that our analytic non-linear mass is a better estimation of the exact numerical solution in Regions I and II. As expected, the point mass approximation generically overestimates the accretion in these regions. In Region III, all three lines essentially overlap and it becomes difficult to distinguish the spherically-symmetric extended loop case from the point-mass approximation. Thus, the point mass is a good assumption for small loops and late accretion onto larger loops, since shells in Region III dominate the turnaround radius in these cases.


In this figure, we illustrate the total mass of the loop by the dotted green line. Once the accreted mass exceeds the loop mass, the existence of the loop is unimportant as the overdensity will continuously accrete matter even if the loop disappears. For our benchmark case, this happens soon after the turnaround shell enters Region II, significantly later than the point-mass case. Furthermore, the time that it takes for the accreted mass to dominate the extended loop mass depends on the size of a given loop, with larger loops taking a longer time. Recall that for the point-mass case, the non-linear mass is proportional to the total mass of the loop\footnote{Here we only consider loops forming in the radiation phase.}. The accreted mass then exceeds the loop mass at a fixed time independent of the loop radius. In the extended source case, only a fraction of the loop mass is inside any given shell at $z>z_{\rm c}^{\rm (II)/(III)}$, so the overall non-linear mass is less than in the point-mass case at early times. Thus, the time for the accreted mass to become dominant is always longer when considering the finite size of loops.

\bfig
	\centering
	\includegraphics[scale=0.55]{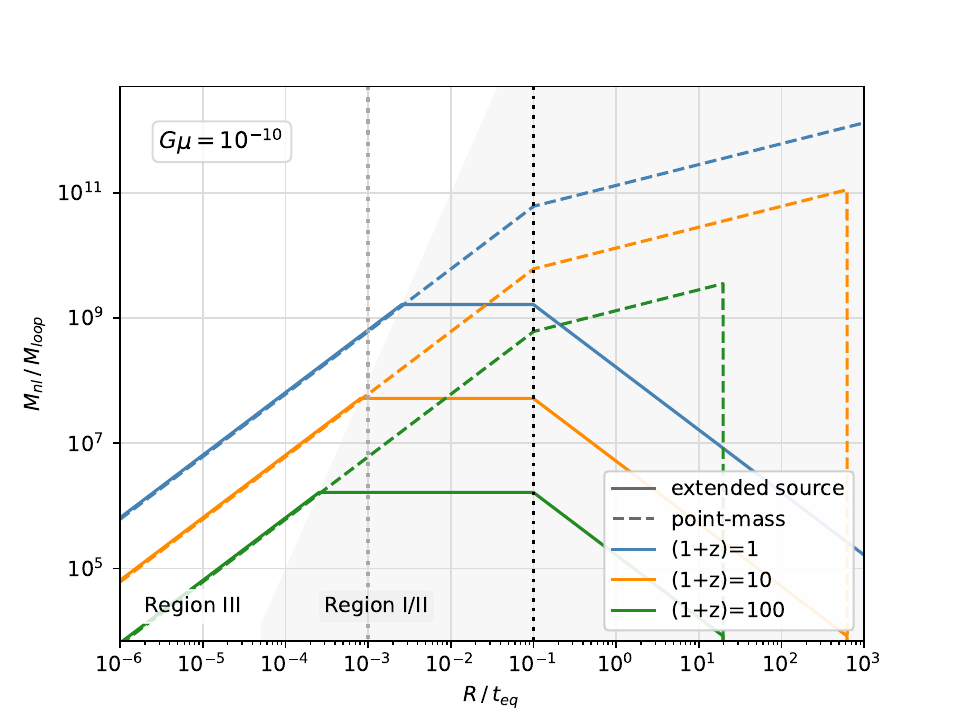}
	\caption{The non-linear mass of an overdensity at different redshift slices as a function of radius, normalized to the mass of the loop itself ($M_{\rm loop} = \beta \mu R_0$). Regions I and II correspond to the light gray region whereas Region III is the white region in this figure. The vertical dotted black line shows the radius of loops which formed at $t_{\rm eq}$, before and after which the initial accretion times have different forms. Additionally, we label our benchmark case  with radius $R_0=10^{-3}t_{\rm eq}$ by the dotted gray line. In Regions I and II, the point mass approximation grossly overestimates the accretion rate, particularly for loops which form in the matter era. The darker dotted line illustrates the characteristic radius of loops forming at $t_{\rm eq}$.}
	\label{fig3-4}
\efig

The non-linear masses as a function of the loop radius are illustrated in Fig \ref{fig3-4}. We show the non-linear mass at redshift slices of $z=1,\ 10,$ and $100$ by the blue, orange, and green lines respectively. The dashed lines represent the case where we view the loops as point masses, and the solid lines correspond to the extended loop case. The two vertical dotted lines denote two typical loop radii. $R_0=10^{-3}t_{\rm eq}$, which is the benchmark value we consider in Fig.~\ref{fig3-2} and \ref{fig3-3}, and $R_0=0.1 t_{\rm eq}$, which is the radius of loop forming at $t_{\rm eq}$. The cutoffs seen at large loop radius are due to the fact that the largest loops at any given time are of size $R_{0,\rm max} = \alpha t$.

The larger the loop, the longer it takes to enter into Region III accretion, yielding greater discrepancies between the point mass and extended loop cases. Indeed this is intuitive, as a larger loop has its mass smeared out over a larger radius when considering the time-averaged density profile. In this figure, we also show the accretion onto loops forming after matter-radiation equality, with $R_0 \geq 0.1 t_{\rm eq}$.

For the extended source case, non-linear masses related to Regions I/II, and III (marked by the light gray and white regions respectively\footnote{Note that we do not distinguish Region I and Region II here as we only consider the leading-order expressions of the non-linear mass. This is according to the discussion in section \ref{ch3-2-2}.}) have different dependencies on the loop radius. In Region III (the left white region), the solid lines almost overlap with the dashed lines and their dependencies on the loop radius are the same, i.e. $M_{\rm nl}\propto R_0$ \footnote{We find in Fig.~\ref{fig3-1} that loops forming in the matter era never enter Region III, so we do not need to discuss this case here.}. While in Region I and II (the light gray region), the accreted non-linear mass $M_{\rm nl}$ is independent of the loop radius for loops forming in the radiation era, yielding the plateau regions. For loops forming in the matter epoch, the initial accretion time is $t_{\rm f}=R_0/\alpha$ instead of $t_{\rm eq}$, leading to the decrease of the non-linear mass over loop radius.

\section{Accretion onto moving loops} \label{ch4}
In the previous section, we ignored the velocity of the loops. However, due to the relativistic speed of long strings, loops are likely to be born with significant velocities \cite{CSsimuls1,CSsimuls2,CSsimuls3,CSsimuls4,CSsimuls5,CSsimuls6,CSsimuls7,CSsimuls8,CSsimuls9,CSsimuls10}. Hence, we should also discuss the accretion onto moving loops in which the assumption of spherical accretion is no longer valid. We still use the Zel'dovich approximation to study the evolution of turnaound shells around a moving loop, resorting now to a description in Cartesian coordinates instead of spherical coordinates. We begin by reviewing the calculation for a moving point mass \cite{movingCSpt1}, and then generalize to an extended loop.

\subsection{Accretion onto a moving point mass} \label{ch4-1}

In Cartesian coordinates, the position and the displacement of a given mass shell are vectors. If we label the comoving coordinate of the loop as $\bbf x^{\rm c}$, the EOM for the vector displacement $\bbf\psi$ with a point-mass source is \cite{CS2,movingCSpt1}.
\ba
\ddot{\bbf\psi}+\frac{4}{3}t^{-1} \dot{\bbf\psi}-\frac23 t^{-2}{\bbf\psi}&=\frac{\beta G\mu R_0(\bbf q - \bbf x^{\rm c})}{|\bbf q - \bbf x^{\rm c}|^3}\left(\frac{t_0}{t}\right)^2.
\end{align}

Here we assume that the loop moves along the z-axis, with an initial (comoving) velocity $v_{\rm i}^{\rm c}$ at the beginning of accretion $t_{\rm i}$, and that the peculiar velocity of the loop will only be redshifted due to the expansion of the universe. Therefore, the loop position can be labelled as $\bbf x^{\rm c}=(0,0,z_{\rm loop}^{\rm c})$, with
\ba
z_{\rm loop}^{\rm c}(t)=3v_{\rm i}^{\rm c}t_{\rm i}[1-(t_{\rm i}/t)^{1/3}]. \label{eq:zloop}
\end{align}
We can define an asymptotic comoving displacement of the loop at $t\gg t_i$, that is $d_{\rm i}^{\rm c}\equiv 3v_{\rm i}^{\rm c}t_{\rm i}$, from Eq. \eqref{eq:zloop} as the factor $(t_{\rm i}/t)$ tends to 0 at late times.

Since this accretion has an azimuthal symmetry around the z-axis, we only need to consider particles in the $x-z$ plane, i.e. we can set $q_y=\psi_y=0$. This assumption will not reduce the universality of our discussion. With this, the EOM of $\psi_x$ and $\psi_z$ becomes
\begin{widetext}
\ba
\ddot\psi_x+\frac{4}{3}t^{-1} \dot\psi_x-\frac23 t^{-2}\psi_x&=\frac{\beta G\mu R_0 q_x}{\bigl[q_x^2+\bigl(q_z-z_{\rm loop}^{\rm c}(t)\bigr)^2\bigr]^{3/2}}\left(\frac{t_0}{t}\right)^2, \label{eq-EOMx}\\
\ddot\psi_z+\frac{4}{3}t^{-1} \dot\psi_z-\frac23 t^{-2}\psi_z&=\frac{\beta G\mu R_0 \bigl[q_z-z_{\rm loop}^{\rm c}(t)\bigr]}{\bigl[q_x^2+\bigl(q_z-z_{\rm loop}^{\rm c}(t)\bigr)^2\bigr]^{3/2}}\left(\frac{t_0}{t}\right)^2. \label{eq-EOMz} 
\end{align}
\end{widetext}

The approximate solution of the above EOM for $t\gg t_{\rm i}$ was studied by Bertschinger \cite{movingCSpt1}:
\ba
\psi_x&=b(t)d_{\rm i}^{\rm c}\left[g_1(\bbf q,a)-g_1(\bbf q,a_{\rm i})\right], \label{movingptsol}\\
\psi_z&=b(t)d_{\rm i}^{\rm c}\left[g_2(\bbf q,a)-g_2(\bbf q,a_{\rm i})\right],
\end{align}
where 
\ba
b(t)&\equiv\frac{1}{15}\frac{\beta G\mu R_0}{v_{\rm i}^{\rm c}{}^3 t_{\rm i}}a_{\rm i}^{-4} a(t),\\
g_1(\bbf q, a)&=\frac{q_x^2+(q_z-d_{\rm i}^{\rm c})[q_z-z_{\rm loop}^{\rm c}(a)]}{q_x R^{\rm c}(\bbf q, a)},\\
g_2(\bbf q, a)&=\frac{d_{\rm i}^{\rm c} (a/a_{\rm i})^{-1/2}}{R^{\rm c}(\bbf q, a)}-\ln\left[R^{\rm c}(\bbf q, a)+q_z-z_{\rm loop}^{\rm c}(a)\right],
\end{align}
with $R^{\rm c}(\bbf q,a)\equiv\sqrt{q_x^2+[q_z-z_{\rm loop}^{\rm c}(a)]^2}$ being the comoving distance between the test particle and the moving loop, and $d_{\rm i}^{\rm c}=3v_{\rm i}^{\rm c}t_{\rm i}$ is the asymptotic comoving displacement of the loop at $t\gg t_{\rm i}$.

As usual, we assume that the turnaround surface satisfies $\dot h_x=0$, i.e. particles in this surface turnaround in the $x$ (and $y$) direction while they may continue to expand in the $z$-direction. In Fig.~\ref{fig4-1}, we compare the turnaround surface sourced by the moving point mass at different times derived from both the approximate analytical displacement in Eq.~\eqref{movingptsol} (dotted lines) and the exact numerical solution of Eq.~\eqref{eq-EOMx} (dashed lines). Different colors correspond to different times denoted in the top left panel. We find that the analytical solution is a good approximation at late times, but in the early stages, it predicts a larger turnaround shell near to the position of the loop. This is because the transient term which is ignored in Eq.~\eqref{eq:zloop} is mildly important when $t\simeq t_{\rm i}$ \cite{movingCSpt1}.

\subsection{Accretion onto an extended moving loop} \label{ch4-2}
\bfs
	\includegraphics[scale=0.6]{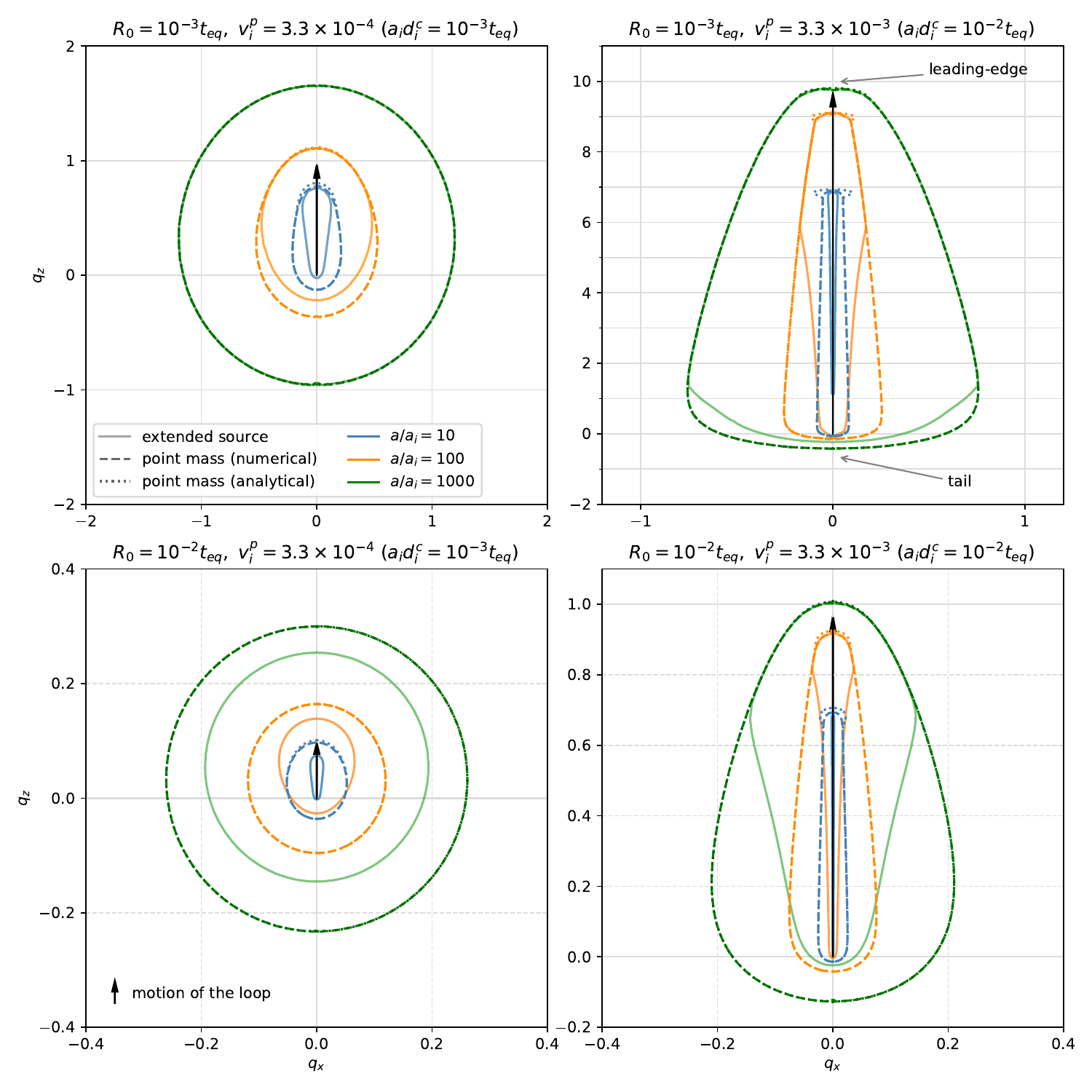}
	\caption{Comparing the turnaround surface sourced by a moving loop when viewed as a point mass (dotted lines for the analytical solution and dashed lines for the numerical result), and as an extended source (solid). All cases take a string tension of $G\mu=10^{-10}$. The blue, orange and green lines correspond to $a/a_{\rm i}=10,\ 100$, and $1000$ respectively. The black arrows show the (comoving) displacement of the loop at $a/a_{\rm i}=1000$ (Eq.~\eqref{eq:zloop}). The coordinates $q_x$ and $q_z$ here are scaled by the comoving loop radius at $t_{\rm eq}$, i.e. $R_0/a(t_{\rm eq})$. Note that the coordinate ranges are different in these panels but the grid spacing of both of them is the the same. \textit{Top panels}: Turnaround surfaces for a moving loop with radius $R_0=10^{-3}t_{\rm eq}$. The two top panels show loops with different (physical) initial velocities $v_{\rm i}^{\rm p}=3.3\times10^{-4}$ (left panel) and $v_{\rm i}^{\rm p}=3.3\times10^{-3}$ (right panel). \tit{Bottom panels}: Accretion onto a moving loop with radius $R_0=10^{-2} t_{\rm eq}$. The initial velocities in the two bottom panels are the same as the corresponding above panels.}
	\label{fig4-1}
\efs

We now study the accretion onto moving extended loops. In the previous sections, we saw that loops forming in the matter era have limited accretion and thus are of less cosmological significance. Since incorporating a finite loop velocity will serve to dilute the accretion efficiency, we choose to specialize our discussion to only those loops which form before matter radiation equality. 

To take into account the finite loop size, we replace the total loop mass $\beta G\mu R_0$ in Eqs. \eqref{eq-EOMx} and \eqref{eq-EOMz} by the enclosed mass $M(aR^{\rm c}(\bbf q,a))$ in Eq.~\eqref{eqenclosedmass},
\ba
M(aR^{\rm c})&=\left\{\begin{array}{ll}
\frac2\pi \beta\mu R_0\arcsin\left(\frac{a R^{\rm c}}{R_0}\right)\ &(aR^{\rm c}<R_0)\\
\beta\mu R_0\ &(aR^{\rm c}\geq R_0)
\end{array}\right.,\nonumber
\end{align}
yielding the EOMs corresponding to a finite-size moving loop. Note that $R^{\rm c}(\bbf q,a)\equiv\sqrt{q_x^2+[q_z-z_{\rm loop}^{\rm c}(a)]^2}$ is the comoving distance between the initial position of the test particle and the center of mass of the moving loop. For $a(t)R^{\rm c}(\bbf q,a(t))<R_0$, the test particle resides inside the loop, and the EOMs become
\begin{widetext}
\ba
\ddot\psi_x+\frac{4}{3}t^{-1} \dot\psi_x-\frac23 t^{-2}\psi_x&=\frac2\pi\frac{\beta (G\mu)R_0 q_x \arcsin\left[aR^{\rm c}(\bbf q,a)/R_0\right]}{R^{\rm c}(\bbf q,a)^3}\left(\frac{t_0}{t}\right)^2,\\
\ddot\psi_z+\frac{4}{3}t^{-1} \dot\psi_z-\frac23 t^{-2}\psi_z&=\frac2\pi\frac{\beta (G\mu)R_0\bigl[q_z-z_{\rm loop}^{\rm c}(a)\bigr]\arcsin\bigl[aR^{\rm c}(\bbf q,a)/R_0\bigr]}{R^{\rm c}(\bbf q,a)^3}\left(\frac{t_0}{t}\right)^2.
\end{align}
\end{widetext}
In the opposite limit ($a(t)R^{\rm c}(\bbf q,a(t))\geq R_0$), the EOMs are the same as Eqs. \eqref{eq-EOMx} and \eqref{eq-EOMz} (i.e. the point mass approximation is valid).

We find no simple analytic form for the solution in the case of an extended loop, so we solve this equation numerically and utilize the above turnaround criterion, $\dot h_x=0$, to compute the turnaround surface for the extended loop case. We compare the turnaround surface for the moving point-mass case and the moving extended loop case in Fig.~\ref{fig4-1}. The solid lines correspond to the extended-loop case whereas the dashed lines and the dotted lines refer to the point-mass scenario predicted by the numerical result of the EOM in \eqref{eq-EOMx} and by the (approximate) analytical solution in Eq.~\eqref{movingptsol}. We show the turnaround surface at three different times $a/a_{\rm i}=10,\ 100,\ 1000$ as the blue, orange and green curves respectively. In this figure, all coordinates are scaled by the comoving loop radius at $t_{\rm eq}$, that is $R_0/a(t_{\rm eq})$. We do not distinguish between the initial accretion time and $t_{\rm eq}$ here since we are only considering loops which formed in the radiation phase.

The velocity of a loop after its creation is in general rather uncertain, with different simulations finding different average values. Loop velocities are also expected to follow a distribution, so in this section we select two typical values for the velocity in order to study the qualitative impact on the geometry of the turnaround surface. 

The top left panel illustrates the turnaround surface for a loop with radius $R_0=10^{-3}t_{\rm eq}$ (our typical benchmark case) and a physical velocity of $v_{\rm i}^{\rm p}=3.3\times 10^{-4}$ at $t_{\rm eq}$. Note that the velocity of a loop decays with the expansion of the universe after it forms. This loop velocity corresponds to an asymptotic displacement of $d_{\rm i}^{\rm c}=10^{-3}t_{\rm eq}/a_{\rm i}=R_0/a_{\rm i}$, implying that the loop will move approximately one comoving radius unit (displayed by the gridlines in the top left panel of Fig.~\ref{fig4-1}) during its lifetime. Note that the physical motion is much larger than the loop radius at late times since the loop does not expand with Hubble flow. In this panel, the turnaround surfaces with respect to the extended loop are smaller than the point-mass case at $a/a_{\rm i}=10$ and 100 as they have smaller ``tails''\footnote{The "tail" and the "leading edge" of the turnaround surface are illustrated in the top right panel of Fig.~\ref{fig4-1}.}. This can be explained by the fact that extended loops accrete less matter in the early stages when the loop is nearer to its original position, exhibited by the fact that the comoving turnaround radius ($q_x$) is smaller in this region. This situation changes by $a/a_{\rm i}=1000$. At these late times, the three turnaround surfaces (green lines) exactly overlap, implying that we enter Region III accretion before $a/a_{\rm i}=1000$. For comparison, a static loop with the same radius enters Region III at $a/a_{\rm i}\approx 520$, which is consistent with the result from the moving loop.

The top right panel shows the accretion onto a loop with the same radius as the left one but with velocity a factor of $10$ greater ($v_{\rm i}^{\rm p}=3.3\times10^{-3}$ and $d_{\rm i}^{\rm c}=10R_0/a_{\rm i}$). Compared to the top left panel, the turnaround surface is much sharper due to the higher velocity, though the qualitative conclusions are the same. The extended loop will lead to smaller tails and the turnaround surface is significantly altered compared to the point mass case at early stages. The deviation becomes much smaller at late times since the turnaround shells tend to be greater than the loop size (similar to Region III accretion in the static case), meaning that the extended density profile of the loop is less important. However, in contrast to the left panel, when $a/a_{\rm i}=1000$, the turnaround surfaces from the point-mass and extended scenarios do not perfectly overlap. This is because the loop moves so fast that turnaround shells do not ever enter Regions II and III before the center of mass of the loop is significantly displaced. Hence, the finite size of the loop is not negligible even at late times. 

The bottom panels in Fig.~\ref{fig4-1} correspond to moving loops with a larger radius ($R_0=10^{-2} t_{\rm eq}$) and the same initial velocities (at $t_{\rm eq}$) as the top panels. 
Compared to the top panels, the turnaround surfaces sourced by moving extended loops in the bottom panels are much smaller than the point-mass case at all of the three times $a/a_{\rm i}=10,\ 100,$ and 1000 (note the difference in scale). This is also consistent with the conclusion we get from static loops, that larger loops are more significantly affected by the smeared-out density profile. Besides, the shapes of these solid turnaround surfaces in the two left/right panels are similar, which also agrees with the conclusion reached in the static loop case, where the shape and size of the turnaround shells are independent of the loop radius when they evolve inside the loop.

We can also compare the two right panels, in which although the turnaround surface associated with the extended loop is still much smaller than the point-mass case, the ``leading edge'' of the surfaces almost overlap with each other at each of our benchmark times. This happens because the displacement of the loop is much greater than the size of the loop, which is roughly constant in physical coordinates when we ignore the decay of the loop and thus, shrinks as $a(t)^{-1}$ in comoving coordinates. Therefore, the accretion at early stages is negligible in the ``leading-edge'' region and does not significantly affect this side of the turnaround surface.

Though this analysis has largely been qualitative, we reach similar conclusions as to the static loop case discussed in Section \ref{ch3}. In the early stages of accretion (Region I/II), the turnaround shells seeded by an extended loop are considerably smaller than the point-mass case, because the source mass inside these shells is only a fraction of the entire loop. However, if the velocity of the loop is large enough, that is if $a_{\rm i}d_{\rm i}^{\rm c}$ is comparable to the loop radius $R_0$ (for example, in the top and bottom right panels of Fig.~\ref{fig4-1}), the leading-edge turnaround surface of the extended loop will reach the point-mass case rapidly and the main difference between the two cases will be the shape of the tail. In the bottom left panel the loop velocity is very small, and the solid curves do not reach the dashed curves in either dimension, but it is qualitatively clear that the leading edge of the surface more quickly approaches that of the point-mass case.

Note that we present two illustrative values of the initial velocity in Fig.~\ref{fig4-1}, which lead to two different shapes of the turnaround surface. In the case of a more relativistic velocity \cite{CS-structure}, the accretion onto a loop would be much more filamentary, and thus, more difficult to illustrate the turnaround surface.

\section{Conclusions and discussion} \label{ch5}
In this article, we have computed the accretion onto cosmic string loops, taking into account their finite size and oscillations, and compared this to the point-mass approximation usually applied in the literature \cite{CS2,movingCSpt1}.

To describe accretion onto an extended loop, we have found it useful to divide the surrounding shells into three distinct regions. Turnaround shells in Regions I and II originate inside the loop, so their evolution will be affected by this finite size. As expected, shells in Region III are well described by the point mass approximation as they originate and turn around outside of $R_0$. 

For static loops, we derived a leading-order power-law expression for the non-linear mass sourced by extended loops in Regions I and II, which we reiterate here,
\ba
M_{\rm nl}^{({\rm I}/{\rm II})} \approx 1.6\times10^9 M_\odot\cdot \left(\frac{G\mu}{10^{-10}}\right)^{3/2}\left(1+z\right)^{-3/2}. \nonumber
\end{align}
An interesting observation is the independence of this expression on the loop radius. This is due to a cancellation of $R_0$ terms when taking the small angle approximation of the time-averaged density profile. In addition, the overall non-linear mass $M_{\rm nl}^{({\rm I}/{\rm II})}$ is smaller than the point-mass case, but grows faster with redshift. This happens because a given turnaround shell feels increasingly more loop mass as it expands outwards with the Hubble flow when it is inside the loop, in contrast to the point mass case where the total source term is fixed with regard to shell expansion. The leading-order solution in Region II inherits some of the properties of the solution in Region I, thus, $M_{\rm nl}^{({\rm I})}$ and $M_{\rm nl}^{({\rm II})}$ have the same redshift and string tension dependence. Deviations between leading-order growth in Regions I and II appear when a given turnaround shell crosses the boundary of the loop ($a(t) q_{\rm nl} \simeq R_0$). These appear as higher-order terms in our expressions so we neglect them. Once a loop reaches Region III accretion, its non-linear mass quickly converges to what is derived in the point mass approximation. The critical redshift between Region II and III accretion is proportional to $R_0^{-2}$ which means that the larger the loop, the later it will reach Region III and the longer the finite loop size will affect the accretion. For $G\mu=10^{-10}$, loops with $R_0 > 2.5\times10^{-3}t_{\rm eq}$ will not enter into Region III before the present time. This includes all loops generated in the matter era. We compare the non-linear masses seeded by loops in the point-mass and extended profile formalisms in Fig.~\ref{fig3-4}.

We perform a more qualitative analysis on extended loops with non-trivial peculiar velocities in Sec.~\ref{ch4}. The impact of loop size and velocity on the accretion dynamics is shown in Fig.~\ref{fig4-1}. From this figure, we see that similar to the static loop case, extended moving loops decrease the size of the turnaround surface in Regions I/II, 
while in Region III, 
the turnaround surface corresponding to an extended source almost overlaps one derived in the point-mass case. The size of a given moving loop exerts a strong influence on the tail of the turnaround surfaces, whereas the leading edges reach the point-mass case much more rapidly, even before $z^{({\rm II}/{\rm III})}_{\rm c}$ if $a_{\rm i}d_{\rm i}^{\rm c}$, is comparable to the loop radius. 

Note that we did not mention the evolution of shells after turnaround. This is because the EOM for $\psi$ is not valid after the turnaround time $t_{\rm ta}$, as the displacement of a mass shell becomes comparable to the initial comoving coordinate ($q$). After turning around, the shells decouple from the Hubble flow, cease their expansion, and begin to collapse. It takes roughly one Hubble time for these shells to virialize and give rise to a uniform temperature in this overdensity, after which the radius of the virialized halo is about $1/4$ of the turnaround radius \cite{Bryce,virial}.

\bfig
	\centering
	\includegraphics[scale=0.55]{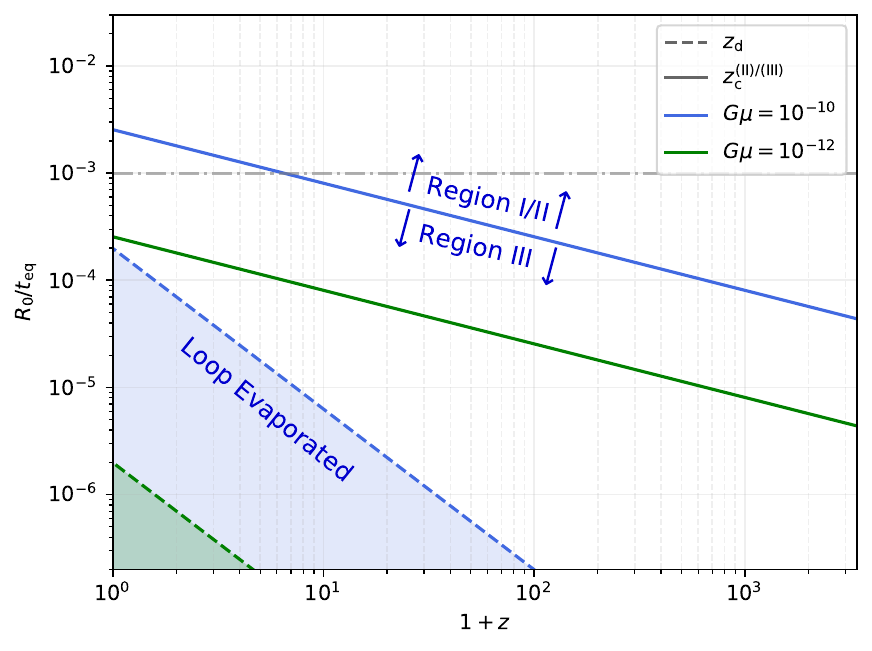}
	\caption{Comparing the decay redshift due to gravitational radiation to the critical redshift between Regions II and III. Dashed lines are the decay time for different values of $G\mu$, and solid lines correspond to the critical redshift $z_{\rm c}^{\rm(II)/(III)}$. We find that for $G\mu<10^{-10}$, which is the current upper limit of string tension, the decay redshift is always much smaller than the critical redshift till now ($z_0=0$), which means that all loops will decay after entering Region III. The horizontal dash-dotted line corresponds to the benchmark value of the loop radius $R_0=10^{-3}t_{\rm eq}$.}
	\label{fig5-1}
\efig

Another point we did not consider here is the decay of cosmic string loops. Oscillating cosmic string loops (with $G\mu\lsim 10^{-10}$) mainly lose energy via gravitational radiation with a power of \cite{CS-GW-1,CS-GW-2}
\ba
P_{GW}\sim\gamma G\mu^2,
\end{align}
where $\gamma\sim100$. Therefore, loops with radius $R_0<\gamma\beta^{-1}G\mu t$ will decay into gravitational radiation within one Hubble time. Nevertheless, for $G\mu<10^{-8}$ (an upper bound much greater than the constraint from NANOGrav \cite{CS-NANO-1,CS-NANO-2,CS-NANO-3}), loops will generically be stable before they enter Region III (see Fig.~\ref{fig5-1}). Therefore, for loops decaying after $t_{\rm eq}$, a given loop will always accrete a non-linear mass $M_{\rm nl} \gtrsim M_{\rm loop}$ before evaporating. Then, the existence of the loop itself is no longer important as the overdensity seeded by the loop will become self-sufficient. On the other hand, if a loop decays in the radiation era, i.e. $R_0<\gamma \beta^{-1}G\mu t_{eq}$, we do not consider it to have sourced any such overdensity.

\section*{Acknowledgement}
\noindent RB wishes to thank the Pauli Center and the Institutes of Theoretical Physics and of Particle- and Astrophysics of the ETH for hospitality. The research of RB and HJ at McGill is supported in part by funds from NSERC and from the Canada Research Chair program. BC is thankful for support from an NSERC-PDF.

\appendix

\section{The Zel'dovich approximation} \label{App-A}

We study the EOM of the dark matter accretion onto a static loop by the Zel'dovich approximation in this appendix, following the analysis in Chapter 11.2 of \cite{CS2}.

We express the initial comoving distance from the centre of a loop to a mass shell by $q$. The comoving displacement of this shell (due to the presence of the loop) is then given by $\psi(q,t)$, meaning that the (physical) height of any given shell can be expressed as:
\ba \label{eq:h}
h(q,t)=a(t)[q-\psi(q,t)],
\end{align}
where we only consider evolution in the matter epoch, $a(t)=(t/t_0)^{2/3}$. In the Zel'dovich approximation the dynamics of this shell is described by the Newtonian equations of motion
\ba \label{eq:NewtEq}
\ddot h=-\frac{\partial \Phi}{\partial h},
\end{align}
where gravitational potential can be determined by the Poisson equation
\ba
\nabla^2\Phi =4\pi G\left[\rho_{\rm 0}^{\rm p}(h,t)+\bar\rho_{{\rm CS}}(h)\right].
\end{align}
Here, $\rho_{\rm 0}^{\rm p}(h,t)$ is the (physical) dark matter density with respect to the scale height $h$, and $\bar\rho_{{\rm CS}}(h)$ is the oscillation averaged cosmic string density profile at $h$. In the matter-dominated era, the dark matter density is related to the perturbation through
\ba
\rho_{\rm 0}^{\rm p}(h,t)&=a(t)^3\rho_{\rm bg}^{\rm p}(t)\frac{q^2}{h^2}\frac{\partial q}{\partial h}\\
&\simeq \rho_{\rm bg}^{\rm p}(t)\left(1+2\frac{\psi}{q}+\frac{\partial \psi}{\partial q}\right),
\end{align}
where $\rho_{\rm bg}^{\rm p}(t)=\frac{3H^2}{8\pi G}$ is the (physical) background density.
In the second line, we only preserve the linear terms of $\psi$. 

We now separate the gravitational potential as $\Phi=\Phi_1+\Phi_2$, with $\Phi_1$ corresponding to the dark matter density and $\Phi_2$ the cosmic string source term. Since both the background density and the time-averaged cosmic string loop density are spherically symmetric, we can exploit spherical symmetry to set $\partial_\theta\Phi=\partial_\phi\Phi=0$.

For $\Phi_1$ (the dark matter term), the Poisson equation is: 
\ba
\nabla_h^2\Phi_1&=4\pi G\rho_0^{\rm p}(h,t)\\
\ \ \frac1{h^2}\frac{\partial}{\partial h}\left( h^2\frac{\partial \Phi_1}{\partial h} \right)&\simeq 4\pi G \rho_{\rm bg}^{\rm p}(t)\left(1+2\frac{\psi}{q}+\frac{\partial \psi}{\partial q}\right).
\end{align}

To determine the evolution of the shell height, Eq.~\eqref{eq:NewtEq}, we need to solve for $\frac{\partial \Phi}{\partial h}$ as a function of $q$,
\ba
\frac{\partial\Phi_1}{\partial h}&=\frac1{h^2}\int dh\, h^2\left[ 4\pi G \rho_{\rm bg}^{\rm p}(t)\left(1+2\frac{\psi}{q}+\frac{\partial \psi}{\partial q}\right) \right]\\
&\simeq \frac{4\pi}{3}aG\rho_{\rm bg}^{\rm p}(q+2\psi)=\frac12 a H^2(q+2\psi).
\end{align}
Note that we set $\Phi_1(h=0)=0$ as a simplification.

For $\Phi_2$, since the source is spherically symmetric, the gravitational potential is determined by the total mass inside the shell $M(h)$:
\ba
&\Phi_2(h)=-\frac{GM}{h}.
\end{align}
Differentiating Eq.~\eqref{eq:h} gives us the last input for the EOM,
\ba
\ddot h&=\ddot aq-(\ddot a\psi+2\dot a \dot\psi+a\ddot\psi).
\end{align}

The final version for the EOM of the shell is
\ba
\ddot h&=-\frac{\partial \Phi}{\partial h},\nn
\ \ \ddot aq-(\ddot a\psi+2\dot a \dot\psi+a\ddot\psi)&\simeq-\frac12 a H^2(q+2\psi)-\frac{GM}{a^2q^2},\nn
\Rightarrow\ \ \ddot\psi+\frac{4}{3}t^{-1} \dot\psi-\frac23 t^{-2}\psi&=\frac{GM}{q^2}\left(\frac{t_0}{t}\right)^2. \label{eq-A12}
\end{align}
Note that we linearize the EOM for the displacement $\psi$, which is based on the assumption that the shell displacement $\psi$ is small compared to the initial coordinate $q$. We can check the validity of this assumption by comparing $\psi$ and $q$ at the turnaround time $t_{\rm ta}$ as $\psi$ keeps increasing before this time. To investigate this, we note that at shell turnaround ($\dot{h} = 0$) we have
\ba
\frac{\dot{h}}{a} = 0 &=\frac{\dot a}{a}(q-\psi)-\dot\psi \nonumber \\ 
&\simeq\frac{2}{3t}(q-\psi)-\dot\psi.
\end{align}
The leading term in $\psi$ is proportional to $(t/t_{\rm i})^{3/2}$ in all Regions (see main text) for both the point-mass case and the extended source case, so we have $\dot\psi\sim\frac{2}{3t}\psi$, and we find the condition
\ba
&\ \ \frac{2}{3t}\left[q-2\psi(q,t_{\rm ta}(q))\right]=0\\
&\ \ \psi(q,t_{\rm ta}(q))=\frac12 q.
\end{align}
To reiterate, $\psi$ starts at 0 at $t=t_{\rm i}$, and then increases monotonically until the turnaround condition, when $\dot{h} = 0$, which implies $\psi (t_{\rm ta}) = q/2$. After the turnaround time, we no longer follow the evolution of the shell and instead assume that virialization occurs afterwards. Therefore, we have $\psi \ll q$ for all relevant timescales in our problem. 


We also apply the Born approximation in this EOM, which neglects the change of the gravitational potential on the r.h.s. of Eq.~\eqref{eq-A12}. We can do this because the constant potential is of order $\cO\left(\psi\right)$, and thus, the evolution of the potential should be $\cO\left(\psi\right)^2$ and can be ignored.

Similarly, the initial conditions of $\psi$ are of the same order of $\psi$ so we can only keep their leading terms, which is of $\cO(\psi)$. This is why we ignore the effect of $\psi$ at the crossing time $t_{\rm cross}$ in Eqs. \eqref{eq:tcross}, \eqref{eq-iniII-1}, and \eqref{eq-iniII-2}.

\section{Turnaround radius without approximation $\arcsin (x)\sim x$} \label{App-B}

We can also derive the analytical solution of the displacement $\psi$ without the small angle approximation, but the corresponding turnaround shell can only be computed numerically.

\subsection{Region I}\label{App-B-1}
Without the small angle approximation $\arcsin (x)\sim x$, the inhomogeneous differential equation of $\psi$ is
\ba
\ddot\psi+\frac{4}{3}\frac{\dot\psi}{t}-\frac23 \frac{\psi}{t^2}=\frac2\pi\frac{\beta G\mu R_0}{q^2}\left(\frac{t_0}{t}\right)^2 \arcsin\left(\frac{q\,t^{2/3}}{R_0 t_0^{2/3}}\right),
\end{align}
for shells in Region I ($a(t_{ta}q<R_0$).

We can solve this EOM by a Green's function method. The full solution of $\psi(t)$ is
\ba
\psi(t)&=\psi_2\int_{t_{\rm i}}^t dt'\frac{\psi_1(t')g(t')}{W(t')}-\psi_1\int_{t_{\rm i}}^t dt'\frac{\psi_2(t')g(t')}{W(t')},
\end{align}
where $\psi_1(t)$ and $\psi_2(t)$ are the two homogeneous solutions of $\psi$:
\ba
\left\{\begin{array}{l}
\psi_1(t)=(t/t_0)^{-1}\\
\psi_2(t)=(t/t_0)^{2/3}
\end{array}\right.,
\end{align}
and $g(t)\equiv\frac2\pi\frac{\beta G\mu R_0}{q^2}\left(\frac{t_0}{t}\right)^2 \arcsin(\frac{q\,t^{2/3}}{R_0 t_0^{2/3}})$ is the r.h.s. of the inhomogeneous equation. $W(t)$ is the wronskian of the above two solutions
\ba
W(t)&=\psi_1\dot\psi_2-\psi_2\dot\psi_1=\frac{5}{3}t_0^{1/3} t^{-4/3}.
\end{align}

Inserting these functions, the displacement $\psi$ becomes
\begin{widetext}
\ba
\psi&=\frac{6}{5\pi}\frac{\beta G\mu R_0}{q^2} t_0^2\, t^{2/3}\int_{t_{\rm i}}^t dt't'^{-5/3}\arcsin\left(\frac{q\,t'^{2/3}}{R_0 t_0^{2/3}}\right) - \frac{6}{5\pi}\frac{\beta G\mu R_0}{q^2} t_0^2\, t^{-1}\int_{t_{\rm i}}^t dt'\arcsin\left(\frac{q\,t'^{2/3}}{R_0 t_0^{2/3}}\right).
\end{align}
\end{widetext}
To calculate this integral, we need to introduce a dimensionless parameter $\xi\equiv\frac{q t'^{2/3}}{R_0 t_0^{3/2}}$.
\begin{widetext}
\ba
\psi&= \frac{9}{5\pi} \frac{\beta G\mu}{q}t^{2/3}t_0^{4/3}\int_{\xi_{\rm i}}^{\xi(t)}d\xi\ \xi^{-2} \arcsin\left(\xi\right) -  \frac{9}{5\pi}\frac{\beta G\mu R_0^{5/2}}{q^{7/2}}t_0^3 t^{-1} \int_{\xi_{\rm i}}^{\xi(t)}d\xi\,\xi^{1/2}\arcsin \xi\nn
&=\frac{1}{5\pi}\frac{\beta G\mu R_0}{q^2} t_0^2\left\{-15\arcsin \xi + \left[9\frac{\xi}{\xi_{\rm i}}+6\left(\frac{\xi_{\rm i}}{\xi}\right)^{3/2}\right]\arcsin\xi_{\rm i}  - \frac{4}{\xi}\sqrt{1-\xi^2}\right.\nn
&\ \ \ \ \ \ \ \ \ \ \ \ \ \ \ \ \ \ \ \ \ \ \  + \frac{4}{\xi_{\rm i}}\left(\frac{\xi_{\rm i}}{\xi}\right)^{3/2}\sqrt{1-\xi_{\rm i}^2} + 9\xi\ln\left(\frac{\xi}{\xi_{\rm i}}\right) - 9\xi\ln\left(\frac{1+\sqrt{1-\xi^2}}{1+\sqrt{1-\xi_{\rm i}^2}}\right) \nn
&\left.\ \ \ \ \ \ \ \ \ \ \ \ \ \ \ \ \ \ \ \ \ \ \  + 4\xi^{-3/2}\left[F\left(\arcsin(\xi^{1/2})\big|-1\right) - F\left(\arcsin(\xi_{\rm i}^{1/2})\big|-1\right)\right] \right\},
\end{align}
\end{widetext}
where $\xi_{\rm i}\equiv \xi(t_{\rm i})=\frac{q t_{\rm i}^{2/3}}{R_0 t_0^{3/2}}$ is the $\xi$ parameter at initial time $t_i$, and $F(\phi|m)$ is the elliptic integral of the first kind.

Since we use parameter $\xi$ to express $q$, the turnaround radius should be determined by $\xi_{\rm nl}$ which satisfies:
\ba
\frac{dh}{dt}=\left.\frac{da}{dt}\frac{d\xi}{da}\frac{dh}{d\xi}\right|_{\xi=\xi_{\rm nl}}=0,
\end{align}
since both $\frac{da}{dt}$ and $\frac{d\xi}{da}=\frac{q}{R_0}$ are positive definite, the equation becomes
\ba
\frac{dh}{d\xi}&=0,\\
\Rightarrow\ \ \frac{R_0}{a}&=\frac{\psi}{\xi}+\frac{d\psi}{d\xi}.
\end{align}
Inserting $\psi$ as a function of $\xi_{\rm i}$ and $a/a_{\rm i}\equiv\xi/\xi_{\rm i}$ (instead of $\xi_{\rm i}$ and $\xi$), the equation becomes
\begin{widetext}
\ba
\frac{5\pi}{\beta G\mu}\left(\frac{R_0}{t_{\rm eq}}\right)^2\frac{z_{\rm i}^{3}}{z_{\rm eq}^3}\cdot\frac{a_{\rm i}}{a}\xi_{\rm i,nl}^2=&-15\frac{\arcsin \left(\frac{a}{a_{\rm i}}\xi_{\rm i,nl}\right)}{\frac{a}{a_{\rm i}}\xi_{\rm i,nl}}-\frac{11}{\sqrt{1-\left(\frac{a}{a_{\rm i}}\xi_{\rm i,nl}\right)^2}} + \left[\frac{18}{\xi_{\rm i,nl}} - \frac{3}{\xi_{\rm i,nl}}\left(\frac{a}{a_{\rm i}}\right)^{-5/2}\right]\arcsin\xi_{\rm i,nl} \nn
&-2\left(\frac{a}{a_{\rm i}}\right)^{-5/2}\xi_{\rm i,nl}^{-2}\sqrt{1-\xi_{\rm i,nl}^2}+9+18\ln\left(\frac{a}{a_{\rm i}}\right) + \frac{9\left(\frac{a}{a_{\rm i}}\xi_{\rm i,nl}\right)^2}{1-\left(\frac{a}{a_{\rm i}}\xi_{\rm i,nl}\right)^2+\sqrt{1-\left(\frac{a}{a_{\rm i}}\xi_{\rm i,nl}\right)^2}}\nn
&-18\ln\left(\frac{1+\sqrt{1-\left(\frac{a}{a_{\rm i}}\xi_{\rm i,nl}\right)^2}}{1+\sqrt{1-\xi_{\rm i,nl}^2}}\right)  + \frac{2}{\left(\frac{a}{a_{\rm i}}\xi_{\rm i,nl}\right)^2\sqrt{1-\left(\frac{a}{a_{\rm i}}\xi_{\rm i,nl}\right)^2}}\nn
&-2\left(\frac{a}{a_{\rm i}}\xi_{\rm i,nl}\right)^{-5/2}\left[F\left(\arcsin\left[(\xi_{\rm i,nl}{a}/{a_{\rm i}})^{1/2}\right]\big|-1\right) - F\left(\arcsin\left(\xi_{\rm i,nl}^{1/2}\right)\big|-1\right)\right].
\end{align}
\end{widetext}

From here, we can solve $\xi_{\rm i,nl}$ as a function of $a/a_{\rm i}$, $G\mu$ and $R_0$ numerically. The result is shown in Fig.\ref{fig3-2}

\subsection{Region II} \label{App-B-2}
In this case, for $t>t_{\rm cross}$, the r.h.s. of the inhomogeneous equation \eqref{eq-A12} becomes $f(t)=\frac{\beta G\mu R_0}{q^2}\left(\frac{t_0}{t}\right)^2$. Due to the different r.h.s., we should integrate the two stages of shells in Region II separately when calculating the comoving displacement $\psi$.
\begin{widetext}
\ba
\psi(t)&=\psi_2\int_{t_{\rm i}}^{t_{\rm cross}} dt'\frac{\psi_1(t')g(t')}{W(t')}+\psi_2\int_{t_{\rm cross}}^{t} dt'\frac{\psi_1(t')f(t')}{W(t')}-\psi_1\int_{t_{\rm i}}^{t_{\rm cross}} dt'\frac{\psi_2(t')g(t')}{W(t')}-\psi_1\int_{t_{\rm cross}}^{t} dt'\frac{\psi_2(t')f(t')}{W(t')}\nn
&= \frac{9}{5\pi} \frac{\beta G\mu}{q}t^{2/3}t_0^{4/3}\Bigg[\int_{\xi_{\rm i}}^{1}d\xi\ \xi^{-2} \arcsin\left(\xi\right)+\frac\pi2\int_{1}^{\xi(t)}d\xi\ \xi^{-2}\Bigg] -  \frac{9}{5\pi}\frac{\beta G\mu R_0^{5/2}}{q^{7/2}}t_0^3 t^{-1} \Bigg[\int_{\xi_{\rm i}}^{1}d\xi\,\xi^{1/2}\arcsin \xi + \int_{t_{\rm cross}}^{t}dt\Bigg]\nn
&=\frac{\beta}{5\pi}\frac{G\mu}{R_0}a_{\rm i}^2\,t_0^{2}\cdot\xi_{\rm i}^{-2}
\Bigg[\left(9\frac{a}{a_{\rm i}}+6\left(\frac{a}{a_{\rm i}}\right)^{-3/2}\right)\arcsin(\xi_{\rm i})-9\xi_{\rm i}\frac{a}{a_{\rm i}}\ln\left(\frac{\xi_{\rm i}}{1+\sqrt{1-\xi_{\rm i}^2}}\right)+\frac{\Gamma\left(\frac14\right)^2}{\sqrt{2\pi}}\left(\frac{a}{a_{\rm i}}\right)^{-3/2}\xi_{\rm i}^{-3/2}\nn
&\ \ \ \ \ \ \ \ \ \ \ \ \ \ \ \ \ \ \ \ \ \ \ \ \ \ \ \ \ \ + 4\left(\frac{a}{a_{\rm i}}\right)^{-3/2}\sqrt{\xi_{\rm i}^{-2}-1} - 4\left(\frac{a}{a_{\rm i}}\right)^{-3/2}\frac1{\xi_{\rm i}} {}_2F_1\left(\frac14,\frac12;\frac54;\xi_{\rm i}^2\right)-\frac{15}{2}\pi\Bigg],
\end{align}
\end{widetext}
where $\Gamma(z)$ is the Euler gamma function, and ${}_2F_1(a,b;c;z)$ is the hypergeometric function.

Then, the equation of the turnaround radius is
\begin{widetext}
\ba
\xi_{\rm i,nl}^3 R_0^2 &= \frac{\beta G\mu}{5\pi}t_{\rm i}^2\times\Bigg\{ \left[18\,\frac{a}{a_{\rm i}}-3\bigg(\frac{a}{a_{\rm i}}\bigg)^{-3/2}\right]\arcsin(\xi_{\rm i,nl})-18\,\frac{a}{a_{\rm i}}\,\xi_{\rm i,nl}\ln\Bigg(\frac{\xi_{\rm i,nl}}{1+\sqrt{1-\xi_{\rm i,nl}^2}}\Bigg) - \frac12\frac{\Gamma\left(\frac14\right)^2}{\sqrt{2\pi}}\bigg(\frac{a}{a_{\rm i}}\bigg)^{-3/2}\xi_{\rm i,nl}^{-3/2}\nn
&\ \ \ \ \ \ \ \ \ \ \ \ \ \ \ \ \ \ \ \ \  - 2\,\bigg(\frac{a}{a_{\rm i}}\bigg)^{-3/2}\xi_{\rm i,nl}^{-1}\sqrt{1-\xi_{\rm i,nl}^{2}} + 2\,\bigg(\frac{a}{a_{\rm i}}\bigg)^{-3/2}\xi_{\rm i,nl}^{-1}\,{}_2F_1\left(\frac14,\frac12,\frac54,\xi_{\rm i,nl}^2\right) -\frac{15}{2}\pi\Bigg\}.
\end{align}
\end{widetext}

\subsection{Region III}\label{App-B-3}
In this case, the solution of $\psi$ is compact
\ba
\psi(t)&=\psi_2\int_{t_{\rm i}}^{t} dt'\frac{\psi_1(t')f(t')}{W(t')}-\psi_1\int_{t_{\rm i}}^{t} dt'\frac{\psi_2(t')f(t')}{W(t')}\nn
&=\frac{3\beta G\mu}{2}R_0^{-1}\xi_{\rm i}^{-2}\frac{t_{\rm i}}{a_{\rm i}}\left[-1+\frac35\left(\frac{a}{a_{\rm i}}\right)+\frac25\left(\frac{a}{a_{\rm i}}\right)^{-3/2}\right],
\end{align}
and we can derive the analytical expression of the turnaround radius parameter $\xi_{\rm i,nl}$
\ba
\xi_{\rm i,nl} = \left(\frac{3\beta G\mu}{2}\right)^{1/2}\left(\frac{R_0}{t_i}\right)^{-1}
\left[-1+\frac65\left(\frac{a}{a_{\rm i}}\right)-\frac15\left(\frac{a}{a_{\rm i}}\right)^{-3/2}\right]^{1/2},
\end{align}
which matches the point-mass case. 


\bibliographystyle{utphys}
\bibliography{AccLoopRefs}

\providecommand{\href}[2]{#2}\begingroup\raggedright\begin{thebibliography}{10}

\bibitem{CS1}
R.~H. Brandenberger, ``{Topological defects and structure formation},''
  \href{http://dx.doi.org/10.1142/S0217751X9400090X}{{\em Int. J. Mod. Phys. A}
  {\bfseries 9} (1994) 2117--2190},
  \href{http://arxiv.org/abs/astro-ph/9310041}{{\ttfamily
  arXiv:astro-ph/9310041}}.

\bibitem{CS2}
A.~Vilenkin and E.~P.~S. Shellard, {\em {Cosmic Strings and Other Topological
  Defects}}.
\newblock Cambridge University Press, 7, 2000.

\bibitem{CS3}
M.~B. Hindmarsh and T.~W.~B. Kibble, ``{Cosmic strings},''
  \href{http://dx.doi.org/10.1088/0034-4885/58/5/001}{{\em Rept. Prog. Phys.}
  {\bfseries 58} (1995) 477--562},
  \href{http://arxiv.org/abs/hep-ph/9411342}{{\ttfamily arXiv:hep-ph/9411342}}.

\bibitem{CS4}
R.~Durrer, M.~Kunz, and A.~Melchiorri, ``{Cosmic structure formation with
  topological defects},''
  \href{http://dx.doi.org/10.1016/S0370-1573(02)00014-5}{{\em Phys. Rept.}
  {\bfseries 364} (2002) 1--81},
  \href{http://arxiv.org/abs/astro-ph/0110348}{{\ttfamily
  arXiv:astro-ph/0110348}}.

\bibitem{Kibble1}
T.~W.~B. Kibble, ``{Phase Transitions in the Early Universe},'' {\em Acta Phys.
  Polon. B} {\bfseries 13} (1982) 723.

\bibitem{Kibble2}
T.~W.~B. Kibble, ``{Some Implications of a Cosmological Phase Transition},''
  \href{http://dx.doi.org/10.1016/0370-1573(80)90091-5}{{\em Phys. Rept.}
  {\bfseries 67} (1980) 183}.

\bibitem{wake1}
J.~Silk and A.~Vilenkin, ``{COSMIC STRINGS AND GALAXY FORMATION},''
  \href{http://dx.doi.org/10.1103/PhysRevLett.53.1700}{{\em Phys. Rev. Lett.}
  {\bfseries 53} (1984) 1700--1703}.

\bibitem{wake2}
M.~J. Rees, ``{Baryon concentrations in string wakes at $z\gtrsim 200$:
  implications for galaxy formation and large-scale structure},''
  \href{http://dx.doi.org/10.1093/mnras/222.1.27p}{{\em Phys. Rev. Lett.}
  {\bfseries 57} (1986) 1655--1660}.

\bibitem{wake3}
A.~Stebbins, S.~Veeraraghavan, R.~H. Brandenberger, J.~Silk, and N.~Turok,
  ``{Cosmic String Wakes},'' \href{http://dx.doi.org/10.1086/165697}{{\em
  Astrophys. J.} {\bfseries 322} (1987) 1--19}.

\bibitem{early1}
A.~Vilenkin, ``{Cosmological Density Fluctuations Produced by Vacuum
  Strings},'' \href{http://dx.doi.org/10.1103/PhysRevLett.46.1496}{{\em Phys.
  Rev. Lett.} {\bfseries 46} (1981) 1169--1172}. [Erratum: Phys.Rev.Lett. 46,
  1496 (1981)].

\bibitem{early2}
N.~Turok and R.~H. Brandenberger, ``{Cosmic Strings and the Formation of
  Galaxies and Clusters of Galaxies},''
  \href{http://dx.doi.org/10.1103/PhysRevD.33.2175}{{\em Phys. Rev. D}
  {\bfseries 33} (1986) 2175}.

\bibitem{early3}
H.~Sato, ``Galaxy formation by cosmic strings,'' {\em Progress of theoretical
  physics} {\bfseries 75} no.~6, (1986) 1342--1350.

\bibitem{noacoustic1}
U.-L. Pen, U.~Seljak, and N.~Turok, ``{Power spectra in global defect theories
  of cosmic structure formation},''
  \href{http://dx.doi.org/10.1103/PhysRevLett.79.1611}{{\em Phys. Rev. Lett.}
  {\bfseries 79} (1997) 1611--1614},
  \href{http://arxiv.org/abs/astro-ph/9704165}{{\ttfamily
  arXiv:astro-ph/9704165}}.

\bibitem{noacoustic2}
J.~Magueijo, A.~Albrecht, D.~Coulson, and P.~Ferreira, ``{Doppler peaks from
  active perturbations},''
  \href{http://dx.doi.org/10.1103/PhysRevLett.76.2617}{{\em Phys. Rev. Lett.}
  {\bfseries 76} (1996) 2617--2620},
  \href{http://arxiv.org/abs/astro-ph/9511042}{{\ttfamily
  arXiv:astro-ph/9511042}}.

\bibitem{CS-CMB-1}
T.~Charnock, A.~Avgoustidis, E.~J. Copeland, and A.~Moss, ``{CMB constraints on
  cosmic strings and superstrings},''
  \href{http://dx.doi.org/10.1103/PhysRevD.93.123503}{{\em Phys. Rev. D}
  {\bfseries 93} no.~12, (2016) 123503},
  \href{http://arxiv.org/abs/1603.01275}{{\ttfamily arXiv:1603.01275
  [astro-ph.CO]}}.

\bibitem{CS-CMB-2}
C.~Dvorkin, M.~Wyman, and W.~Hu, ``{Cosmic String constraints from WMAP and the
  South Pole Telescope},''
  \href{http://dx.doi.org/10.1103/PhysRevD.84.123519}{{\em Phys. Rev. D}
  {\bfseries 84} (2011) 123519},
  \href{http://arxiv.org/abs/1109.4947}{{\ttfamily arXiv:1109.4947
  [astro-ph.CO]}}.

\bibitem{CS-CMB-3}
{\bfseries Planck} Collaboration, P.~A.~R. Ade {\em et~al.}, ``{Planck 2013
  results. XXV. Searches for cosmic strings and other topological defects},''
  \href{http://dx.doi.org/10.1051/0004-6361/201321621}{{\em Astron. Astrophys.}
  {\bfseries 571} (2014) A25}, \href{http://arxiv.org/abs/1303.5085}{{\ttfamily
  arXiv:1303.5085 [astro-ph.CO]}}.

\bibitem{RHBrev}
R.~H. Brandenberger, ``{Searching for Cosmic Strings in New Observational
  Windows},'' \href{http://dx.doi.org/10.1016/j.nuclphysbps.2013.10.064}{{\em
  Nucl. Phys. B Proc. Suppl.} {\bfseries 246-247} (2014) 45--57},
  \href{http://arxiv.org/abs/1301.2856}{{\ttfamily arXiv:1301.2856
  [astro-ph.CO]}}.

\bibitem{CS-CMBaniso-1}
N.~Kaiser and A.~Stebbins, ``{Microwave Anisotropy Due to Cosmic Strings},''
  \href{http://dx.doi.org/10.1038/310391a0}{{\em Nature} {\bfseries 310} (1984)
  391--393}.

\bibitem{CS-CMBaniso-2}
R.~Moessner, L.~Perivolaropoulos, and R.~H. Brandenberger, ``{A Cosmic string
  specific signature on the cosmic microwave background},''
  \href{http://dx.doi.org/10.1086/173992}{{\em Astrophys. J.} {\bfseries 425}
  (1994) 365--371}, \href{http://arxiv.org/abs/astro-ph/9310001}{{\ttfamily
  arXiv:astro-ph/9310001}}.

\bibitem{CS-CMB-4}
L.~Hergt, A.~Amara, R.~Brandenberger, T.~Kacprzak, and A.~Refregier,
  ``{Searching for Cosmic Strings in CMB Anisotropy Maps using Wavelets and
  Curvelets},'' \href{http://dx.doi.org/10.1088/1475-7516/2017/06/004}{{\em
  JCAP} {\bfseries 06} (2017) 004},
  \href{http://arxiv.org/abs/1608.00004}{{\ttfamily arXiv:1608.00004
  [astro-ph.CO]}}.

\bibitem{CS-CMB-5}
J.~D. McEwen, S.~M. Feeney, H.~V. Peiris, Y.~Wiaux, C.~Ringeval, and F.~R.
  Bouchet, ``{Wavelet-Bayesian inference of cosmic strings embedded in the
  cosmic microwave background},''
  \href{http://dx.doi.org/10.1093/mnras/stx2268}{{\em Mon. Not. Roy. Astron.
  Soc.} {\bfseries 472} no.~4, (2017) 4081--4098},
  \href{http://arxiv.org/abs/1611.10347}{{\ttfamily arXiv:1611.10347
  [astro-ph.IM]}}.

\bibitem{CS-CMB-6}
R.~Ciuca and O.~F. Hern\'andez, ``{A Bayesian Framework for Cosmic String
  Searches in CMB Maps},''
  \href{http://dx.doi.org/10.1088/1475-7516/2017/08/028}{{\em JCAP} {\bfseries
  08} (2017) 028}, \href{http://arxiv.org/abs/1706.04131}{{\ttfamily
  arXiv:1706.04131 [astro-ph.CO]}}.

\bibitem{CS-CMB-7}
R.~Ciuca, O.~F. Hern\'andez, and M.~Wolman, ``{A Convolutional Neural Network
  For Cosmic String Detection in CMB Temperature Maps},''
  \href{http://dx.doi.org/10.1093/mnras/stz491}{{\em Mon. Not. Roy. Astron.
  Soc.} {\bfseries 485} (2019) 1377},
  \href{http://arxiv.org/abs/1708.08878}{{\ttfamily arXiv:1708.08878
  [astro-ph.CO]}}.

\bibitem{Holder1}
R.~J. Danos, R.~H. Brandenberger, and G.~Holder, ``{A Signature of Cosmic
  Strings Wakes in the CMB Polarization},''
  \href{http://dx.doi.org/10.1103/PhysRevD.82.023513}{{\em Phys. Rev. D}
  {\bfseries 82} (2010) 023513},
  \href{http://arxiv.org/abs/1003.0905}{{\ttfamily arXiv:1003.0905
  [astro-ph.CO]}}.

\bibitem{Holder2}
R.~H. Brandenberger, R.~J. Danos, O.~F. Hernandez, and G.~P. Holder, ``{The 21
  cm Signature of Cosmic String Wakes},''
  \href{http://dx.doi.org/10.1088/1475-7516/2010/12/028}{{\em JCAP} {\bfseries
  12} (2010) 028}, \href{http://arxiv.org/abs/1006.2514}{{\ttfamily
  arXiv:1006.2514 [astro-ph.CO]}}.

\bibitem{wake-LSS}
D.~Maibach, R.~Brandenberger, D.~Crichton, and A.~Refregier, ``{Extracting the
  signal of cosmic string wakes from 21-cm observations},''
  \href{http://dx.doi.org/10.1103/PhysRevD.104.123535}{{\em Phys. Rev. D}
  {\bfseries 104} no.~12, (2021) 123535},
  \href{http://arxiv.org/abs/2107.07289}{{\ttfamily arXiv:2107.07289
  [astro-ph.CO]}}.

\bibitem{Disrael}
D.~C. Neves~da Cunha, J.~Harnois-Deraps, R.~Brandenberger, A.~Amara, and
  A.~Refregier, ``{Dark Matter Distribution Induced by a Cosmic String Wake in
  the Nonlinear Regime},''
  \href{http://dx.doi.org/10.1103/PhysRevD.98.083015}{{\em Phys. Rev. D}
  {\bfseries 98} no.~8, (2018) 083015},
  \href{http://arxiv.org/abs/1804.00083}{{\ttfamily arXiv:1804.00083
  [astro-ph.CO]}}.

\bibitem{Hannah}
M.~Blamart, H.~Fronenberg, and R.~Brandenberger, ``{Signal of cosmic strings in
  cross-correlation of 21-cm redshift and CMB polarization maps},''
  \href{http://dx.doi.org/10.1088/1475-7516/2022/11/012}{{\em JCAP} {\bfseries
  11} (2022) 012}, \href{http://arxiv.org/abs/2205.02725}{{\ttfamily
  arXiv:2205.02725 [astro-ph.CO]}}.

\bibitem{NANOgrav1}
J.~J. Blanco-Pillado, K.~D. Olum, and X.~Siemens, ``{New limits on cosmic
  strings from gravitational wave observation},''
  \href{http://dx.doi.org/10.1016/j.physletb.2018.01.050}{{\em Phys. Lett. B}
  {\bfseries 778} (2018) 392--396},
  \href{http://arxiv.org/abs/1709.02434}{{\ttfamily arXiv:1709.02434
  [astro-ph.CO]}}.

\bibitem{NANOgrav2}
{\bfseries NANOGrav} Collaboration, Z.~Arzoumanian {\em et~al.}, ``{The
  NANOGrav 12.5 yr Data Set: Search for an Isotropic Stochastic
  Gravitational-wave Background},''
  \href{http://dx.doi.org/10.3847/2041-8213/abd401}{{\em Astrophys. J. Lett.}
  {\bfseries 905} no.~2, (2020) L34},
  \href{http://arxiv.org/abs/2009.04496}{{\ttfamily arXiv:2009.04496
  [astro-ph.HE]}}.

\bibitem{CS-NANO-2}
J.~J. Blanco-Pillado, K.~D. Olum, and J.~M. Wachter, ``{Comparison of cosmic
  string and superstring models to NANOGrav 12.5-year results},''
  \href{http://dx.doi.org/10.1103/PhysRevD.103.103512}{{\em Phys. Rev. D}
  {\bfseries 103} no.~10, (2021) 103512},
  \href{http://arxiv.org/abs/2102.08194}{{\ttfamily arXiv:2102.08194
  [astro-ph.CO]}}.

\bibitem{CS-NANO-1}
{\bfseries NANOGRAV} Collaboration, Z.~Arzoumanian {\em et~al.}, ``{The
  NANOGrav 11-year Data Set: Pulsar-timing Constraints On The Stochastic
  Gravitational-wave Background},''
  \href{http://dx.doi.org/10.3847/1538-4357/aabd3b}{{\em Astrophys. J.}
  {\bfseries 859} no.~1, (2018) 47},
  \href{http://arxiv.org/abs/1801.02617}{{\ttfamily arXiv:1801.02617
  [astro-ph.HE]}}.

\bibitem{CS-NANO-3}
J.~Ellis and M.~Lewicki, ``{Cosmic String Interpretation of NANOGrav Pulsar
  Timing Data},'' \href{http://dx.doi.org/10.1103/PhysRevLett.126.041304}{{\em
  Phys. Rev. Lett.} {\bfseries 126} no.~4, (2021) 041304},
  \href{http://arxiv.org/abs/2009.06555}{{\ttfamily arXiv:2009.06555
  [astro-ph.CO]}}.

\bibitem{NANOGrav2023det}
{\bfseries NANOGrav} Collaboration, G.~Agazie {\em et~al.}, ``{The NANOGrav 15
  yr Data Set: Evidence for a Gravitational-wave Background},''
  \href{http://dx.doi.org/10.3847/2041-8213/acdac6}{{\em Astrophys. J. Lett.}
  {\bfseries 951} no.~1, (2023) L8},
  \href{http://arxiv.org/abs/2306.16213}{{\ttfamily arXiv:2306.16213
  [astro-ph.HE]}}.

\bibitem{Ellis2020}
J.~Ellis and M.~Lewicki, ``{Cosmic String Interpretation of NANOGrav Pulsar
  Timing Data},'' \href{http://dx.doi.org/10.1103/PhysRevLett.126.041304}{{\em
  Phys. Rev. Lett.} {\bfseries 126} no.~4, (2021) 041304},
  \href{http://arxiv.org/abs/2009.06555}{{\ttfamily arXiv:2009.06555
  [astro-ph.CO]}}.

\bibitem{Wang2023}
Z.~{Wang}, L.~{Lei}, H.~{Jiao}, L.~{Feng}, and Y.-Z. {Fan}, ``{The nanohertz
  stochastic gravitational-wave background from cosmic string Loops and the
  abundant high redshift massive galaxies},''
  \href{http://dx.doi.org/10.48550/arXiv.2306.17150}{{\em arXiv e-prints}
  (June, 2023) arXiv:2306.17150},
  \href{http://arxiv.org/abs/2306.17150}{{\ttfamily arXiv:2306.17150
  [astro-ph.HE]}}.

\bibitem{NANOGrav2023}
{\bfseries NANOGrav} Collaboration, A.~Afzal {\em et~al.}, ``{The NANOGrav 15
  yr Data Set: Search for Signals from New Physics},''
  \href{http://dx.doi.org/10.3847/2041-8213/acdc91}{{\em Astrophys. J. Lett.}
  {\bfseries 951} no.~1, (2023) L11},
  \href{http://arxiv.org/abs/2306.16219}{{\ttfamily arXiv:2306.16219
  [astro-ph.HE]}}.

\bibitem{Jerome}
S.~F. Bramberger, R.~H. Brandenberger, P.~Jreidini, and J.~Quintin, ``{Cosmic
  String Loops as the Seeds of Super-Massive Black Holes},''
  \href{http://dx.doi.org/10.1088/1475-7516/2015/06/007}{{\em JCAP} {\bfseries
  06} (2015) 007}, \href{http://arxiv.org/abs/1503.02317}{{\ttfamily
  arXiv:1503.02317 [astro-ph.CO]}}.

\bibitem{Bryce}
B.~Cyr, H.~Jiao, and R.~Brandenberger, ``{Massive black holes at high redshifts
  from superconducting cosmic strings},''
  \href{http://dx.doi.org/10.1093/mnras/stac1939}{{\em Mon. Not. Roy. Astron.
  Soc.} {\bfseries 517} no.~2, (2022) 2221--2230},
  \href{http://arxiv.org/abs/2202.01799}{{\ttfamily arXiv:2202.01799
  [astro-ph.CO]}}.

\bibitem{Bryce2023}
B.~Cyr, J.~Chluba, and S.~K. Acharya, ``{Constraints on the spectral signatures
  of superconducting cosmic strings},''
  \href{http://arxiv.org/abs/2305.09816}{{\ttfamily arXiv:2305.09816
  [astro-ph.CO]}}.

\bibitem{Cyr2023RSB}
B.~{Cyr}, J.~{Chluba}, and S.~K. {Acharya}, ``{A cosmic string solution to the
  radio synchrotron background},''
  \href{http://dx.doi.org/10.48550/arXiv.2308.03512}{{\em arXiv e-prints}
  (Aug., 2023) arXiv:2308.03512},
  \href{http://arxiv.org/abs/2308.03512}{{\ttfamily arXiv:2308.03512
  [astro-ph.CO]}}.

\bibitem{Fixsen2009}
D.~J. Fixsen {\em et~al.}, ``{ARCADE 2 Measurement of the Extra-Galactic Sky
  Temperature at 3-90 GHz},''
  \href{http://dx.doi.org/10.1088/0004-637X/734/1/5}{{\em Astrophys. J.}
  {\bfseries 734} (2011) 5}, \href{http://arxiv.org/abs/0901.0555}{{\ttfamily
  arXiv:0901.0555 [astro-ph.CO]}}.

\bibitem{Dowell2018}
J.~Dowell and G.~B. Taylor, ``{The Radio Background Below 100 MHz},''
  \href{http://dx.doi.org/10.3847/2041-8213/aabf86}{{\em Astrophys. J. Lett.}
  {\bfseries 858} no.~1, (2018) L9},
  \href{http://arxiv.org/abs/1804.08581}{{\ttfamily arXiv:1804.08581
  [astro-ph.CO]}}.

\bibitem{Singal2022}
J.~Singal {\em et~al.}, ``{The Second Radio Synchrotron Background Workshop:
  Conference Summary and Report},''
  \href{http://dx.doi.org/10.1088/1538-3873/acbdbf}{{\em Publ. Astron. Soc.
  Pac.} {\bfseries 135} no.~1045, (2023) 036001},
  \href{http://arxiv.org/abs/2211.16547}{{\ttfamily arXiv:2211.16547
  [astro-ph.CO]}}.

\bibitem{Cowie2023}
F.~J. Cowie, A.~R. Offringa, B.~K. Gehlot, J.~Singal, S.~Heston, S.~Horiuchi,
  and D.~M. Lucero, ``{Diffuse sources, clustering, and the excess anisotropy
  of the radio synchrotron background},''
  \href{http://dx.doi.org/10.1093/mnras/stad1671}{{\em Mon. Not. Roy. Astron.
  Soc.} {\bfseries 523} no.~4, (2023) 5034--5046},
  \href{http://arxiv.org/abs/2306.00829}{{\ttfamily arXiv:2306.00829
  [astro-ph.CO]}}.

\bibitem{CSsimuls1}
A.~Albrecht and N.~Turok, ``{Evolution of Cosmic Strings},''
  \href{http://dx.doi.org/10.1103/PhysRevLett.54.1868}{{\em Phys. Rev. Lett.}
  {\bfseries 54} (1985) 1868--1871}.

\bibitem{CSsimuls2}
D.~P. Bennett and F.~R. Bouchet, ``{Evidence for a Scaling Solution in Cosmic
  String Evolution},'' \href{http://dx.doi.org/10.1103/PhysRevLett.60.257}{{\em
  Phys. Rev. Lett.} {\bfseries 60} (1988) 257}.

\bibitem{CSsimuls3}
B.~Allen and E.~P.~S. Shellard, ``{Cosmic string evolution: a numerical
  simulation},'' \href{http://dx.doi.org/10.1103/PhysRevLett.64.119}{{\em Phys.
  Rev. Lett.} {\bfseries 64} (1990) 119--122}.

\bibitem{CSsimuls4}
V.~Vanchurin, K.~D. Olum, and A.~Vilenkin, ``{Scaling of cosmic string
  loops},'' \href{http://dx.doi.org/10.1103/PhysRevD.74.063527}{{\em Phys. Rev.
  D} {\bfseries 74} (2006) 063527},
  \href{http://arxiv.org/abs/gr-qc/0511159}{{\ttfamily arXiv:gr-qc/0511159}}.

\bibitem{CSsimuls5}
C.~Ringeval, M.~Sakellariadou, and F.~Bouchet, ``{Cosmological evolution of
  cosmic string loops},''
  \href{http://dx.doi.org/10.1088/1475-7516/2007/02/023}{{\em JCAP} {\bfseries
  02} (2007) 023}, \href{http://arxiv.org/abs/astro-ph/0511646}{{\ttfamily
  arXiv:astro-ph/0511646}}.

\bibitem{CSsimuls6}
L.~Lorenz, C.~Ringeval, and M.~Sakellariadou, ``{Cosmic string loop
  distribution on all length scales and at any redshift},''
  \href{http://dx.doi.org/10.1088/1475-7516/2010/10/003}{{\em JCAP} {\bfseries
  10} (2010) 003}, \href{http://arxiv.org/abs/1006.0931}{{\ttfamily
  arXiv:1006.0931 [astro-ph.CO]}}.

\bibitem{CSsimuls7}
J.~J. Blanco-Pillado, K.~D. Olum, and B.~Shlaer, ``{Large parallel cosmic
  string simulations: New results on loop production},''
  \href{http://dx.doi.org/10.1103/PhysRevD.83.083514}{{\em Phys. Rev. D}
  {\bfseries 83} (2011) 083514},
  \href{http://arxiv.org/abs/1101.5173}{{\ttfamily arXiv:1101.5173
  [astro-ph.CO]}}.

\bibitem{CSsimuls8}
J.~J. Blanco-Pillado, K.~D. Olum, and B.~Shlaer, ``{The number of cosmic string
  loops},'' \href{http://dx.doi.org/10.1103/PhysRevD.89.023512}{{\em Phys. Rev.
  D} {\bfseries 89} no.~2, (2014) 023512},
  \href{http://arxiv.org/abs/1309.6637}{{\ttfamily arXiv:1309.6637
  [astro-ph.CO]}}.

\bibitem{CSsimuls9}
P.~Auclair, C.~Ringeval, M.~Sakellariadou, and D.~Steer, ``{Cosmic string loop
  production functions},''
  \href{http://dx.doi.org/10.1088/1475-7516/2019/06/015}{{\em JCAP} {\bfseries
  06} (2019) 015}, \href{http://arxiv.org/abs/1903.06685}{{\ttfamily
  arXiv:1903.06685 [astro-ph.CO]}}.

\bibitem{CSsimuls10}
J.~J. Blanco-Pillado and K.~D. Olum, ``{Direct determination of cosmic string
  loop density from simulations},''
  \href{http://dx.doi.org/10.1103/PhysRevD.101.103018}{{\em Phys. Rev. D}
  {\bfseries 101} no.~10, (2020) 103018},
  \href{http://arxiv.org/abs/1912.10017}{{\ttfamily arXiv:1912.10017
  [astro-ph.CO]}}.

\bibitem{IMBH}
R.~Brandenberger, B.~Cyr, and H.~Jiao, ``{Intermediate mass black hole seeds
  from cosmic string loops},''
  \href{http://dx.doi.org/10.1103/PhysRevD.104.123501}{{\em Phys. Rev. D}
  {\bfseries 104} no.~12, (2021) 123501},
  \href{http://arxiv.org/abs/2103.14057}{{\ttfamily arXiv:2103.14057
  [astro-ph.CO]}}.

\bibitem{JH-JWST}
H.~Jiao, R.~Brandenberger, and A.~Refregier, ``{Early Structure Formation from
  Cosmic String Loops in Light of Early JWST Observations},''
  \href{http://arxiv.org/abs/2304.06429}{{\ttfamily arXiv:2304.06429
  [astro-ph.CO]}}.

\bibitem{CSpt2}
M.~Pagano and R.~Brandenberger, ``{The 21cm Signature of a Cosmic String
  Loop},'' \href{http://dx.doi.org/10.1088/1475-7516/2012/05/014}{{\em JCAP}
  {\bfseries 05} (2012) 014}, \href{http://arxiv.org/abs/1201.5695}{{\ttfamily
  arXiv:1201.5695 [astro-ph.CO]}}.

\bibitem{movingCSpt1}
E.~Bertschinger, ``Cosmological accretion wakes,'' {\em The Astrophysical
  Journal} {\bfseries 316} (1987) 489--496.

\bibitem{CS-structure}
B.~Shlaer, A.~Vilenkin, and A.~Loeb, ``{Early structure formation from cosmic
  string loops},'' \href{http://dx.doi.org/10.1088/1475-7516/2012/05/026}{{\em
  JCAP} {\bfseries 05} (2012) 026},
  \href{http://arxiv.org/abs/1202.1346}{{\ttfamily arXiv:1202.1346
  [astro-ph.CO]}}.

\bibitem{one-scale-1}
E.~J. Copeland, T.~W.~B. Kibble, and D.~Austin, ``{Scaling solutions in cosmic
  string networks},'' \href{http://dx.doi.org/10.1103/PhysRevD.45.R1000}{{\em
  Phys. Rev. D} {\bfseries 45} (1992) 1000--1004}.

\bibitem{one-scale-2}
L.~Perivolaropoulos, ``{COBE versus cosmic strings: An Analytical model},''
  \href{http://dx.doi.org/10.1016/0370-2693(93)91825-8}{{\em Phys. Lett. B}
  {\bfseries 298} (1993) 305--311},
  \href{http://arxiv.org/abs/hep-ph/9208247}{{\ttfamily arXiv:hep-ph/9208247}}.

\bibitem{one-scale-3}
D.~Austin, E.~J. Copeland, and T.~W.~B. Kibble, ``{Evolution of cosmic string
  configurations},'' \href{http://dx.doi.org/10.1103/PhysRevD.48.5594}{{\em
  Phys. Rev. D} {\bfseries 48} (1993) 5594--5627},
  \href{http://arxiv.org/abs/hep-ph/9307325}{{\ttfamily arXiv:hep-ph/9307325}}.

\bibitem{Zel'dovich}
Y.~B. Zeldovich, ``{Gravitational instability: An Approximate theory for large
  density perturbations},'' {\em Astron. Astrophys.} {\bfseries 5} (1970)
  84--89.

\bibitem{virial}
H.~Mo, F.~Van~den Bosch, and S.~White, {\em Galaxy formation and evolution}.
\newblock Cambridge University Press, 2010.

\bibitem{CS-GW-1}
T.~Vachaspati and A.~Vilenkin, ``{Gravitational Radiation from Cosmic
  Strings},'' \href{http://dx.doi.org/10.1103/PhysRevD.31.3052}{{\em Phys. Rev.
  D} {\bfseries 31} (1985) 3052}.

\bibitem{CS-GW-2}
R.~L. Davis, ``{Nucleosynthesis Problems for String Models of Galaxy
  Formation},'' \href{http://dx.doi.org/10.1016/0370-2693(85)90762-2}{{\em
  Phys. Lett. B} {\bfseries 161} (1985) 285--288}.

\end{thebibliography}\endgroup

\end{document}